\documentclass[a4paper,fleqn,usenatbib]{mnras}

\usepackage{mathptmx}
\usepackage{mathrsfs}
\usepackage[T1]{fontenc}
\usepackage{ae,aecompl}

\usepackage{graphicx}	% Including figure files
\usepackage{amsmath}	% Advanced maths commands
\usepackage{amssymb}	% Extra maths symbols

\newcommand{\kms}{\,km\,s$^{-1}$}	% per cm-squared

\newcommand {\sm}{\rm\,M$_\odot$}

\newcommand {\vi} {{\it V--I\/}}

\def\tablefoot#1{\par\vspace*{2ex}%
 \parbox{\hsize}{\leftskip0pt\rightskip0pt
 {\noindent\small\textbf{Notes.}~#1\par}}}
\def\tablefootmark#1{$^{#1}$\,\ignorespaces}
\def\tablefoottext#1#2{$^{(#1)}$~#2}

\title[VLT/FORS2 survey of the Phoenix dwarf]{Prolate rotation and metallicity gradient in the transforming dwarf galaxy Phoenix
\thanks{Based on observations made with ESO telescopes at the La Silla Paranal Observatory under programme IDs 083.B-0252(B) and 71.B-0516}
\thanks{A table containing the measured velocities, metallicities, and CaT equivalent widths of all spectroscopic targets will be made available online at the CDS.}
}

\author[N. Kacharov et al.]{
Nikolay Kacharov,$^{1}$\thanks{E-mail: kacharov@mpia.de}
Giuseppina Battaglia,$^{2,3}$
Marina Rejkuba,$^{4,5}$
\newauthor{}
Andrew A. Cole,$^{6}$
Ricardo Carrera,$^{2,3}$
Filippo Fraternali,$^{7}$
Mark I. Wilkinson,$^{8}$
\newauthor{}
Carme G. Gallart,$^{2,3}$
Mike Irwin,$^{9}$
and Eline Tolstoy$^{10}$
\\
$^{1}$Max Planck Institut f\"{u}r Astronomie, K\"{o}nigstuhl 17, 69117 Heidelberg, Germany \\
$^{2}$Instituto de Astrof\'{i}sica de Canarias, calle V\'{i}a L\'{a}ctea s/n, E38205 - La Laguna (Tenerife), Spain \\
$^{3}$Universidad de La Laguna, Dpto. Astrofisica, E-38206 La Laguna, Tenerife, Spain \\
$^{4}$European Southern Observatory, Karl-Schwarzschild-Str. 2, 85748 Garching bei M\"{u}nchen, Germany \\
$^{5}$Excellence Cluster Universe, Boltzmannstr. 2, 85748, Garching bei M\"{u}nchen, Germany \\
$^{6}$School of Physical Sciences, University of Tasmania, Private Bag 37, Hobart, 7001 TAS, Australia \\
$^{7}$Department of Astronomy, University of Bologna, via Ranzani 1, 40127 Bologna, Italy \\
$^{8}$Department of Physics and Astronomy, University of Leicester, University Road, Leicester LE1 7RH, UK \\
$^{9}$Institute of Astronomy, Madingley Road, Cambridge CB3 OHA, UK \\
$^{10}$Kapteyn Astronomical Institute, University of Groningen, Landleven 12, 9747AD Groningen, Netherlands
}

\date{Accepted 02 Dec. 2016; Received 25 Nov. 2016; in original form 16 Oct. 2016}

\pubyear{2016}

\begin{document}
\label{firstpage}
\pagerange{\pageref{firstpage}--\pageref{lastpage}}
\maketitle

\begin{abstract}
  Transition type dwarf galaxies are thought to be systems undergoing the process of transformation from a star-forming into a passively evolving dwarf, which makes them particularly suitable to study evolutionary processes driving the existence of different dwarf morphological types. 
  Here we present results from a spectroscopic survey of $\sim$200 individual red giant branch stars in the
  Phoenix dwarf, the closest transition type with a comparable luminosity to ``classical'' dwarf galaxies.
  We measure a systemic heliocentric velocity $V_{helio} = -21.2\pm1.0$\kms.~%, confirming the physical association of Phoenix stellar component to the nearby HI cloud at similar velocity.
  Our survey reveals the clear presence of prolate rotation, which is aligned with the peculiar spatial distribution of the youngest stars in Phoenix.
  We speculate that both features might have arisen from the same event, possibly an accretion of a smaller system. 
  The evolved stellar population of Phoenix is relatively metal-poor ($<{\rm[Fe/H]}> = -1.49\pm0.04$\,dex) and shows a large metallicity spread ($\sigma_{\rm [Fe/H]} = 0.51\pm0.04$\,dex), with a pronounced
  metallicity gradient of $-0.13\pm0.01$\,dex per arcmin similar to luminous, passive dwarf galaxies.
  We also report a discovery of an extremely metal-poor star candidate in Phoenix and discuss the importance of correcting
  for spatial sampling when interpreting the chemical properties of galaxies with metallicity gradients.
  This study presents a major leap forward in our knowledge of the internal kinematics of the Phoenix transition type dwarf galaxy, and the first wide area spectroscopic survey of its metallicity 
  properties.
\end{abstract}


%%%%%%%%%%%%%%%%%%%%%%%%%%%%%%%%%%%%%%%%%%%%%%%%%%

%%%%%%%%%%%%%%%%% BODY OF PAPER %%%%%%%%%%%%%%%%%%

\section{Introduction}

Understanding the properties of dwarf galaxies is important not only to put them in their proper cosmological context, 
but also to understand the formation and evolution of the most common type of galaxies. 

%As for larger systems, identifying similarities and
%differences in the physical properties of the various morphological types of dwarf galaxies at present day provides
%insights into the mechanisms that have shaped this galaxy population. 

Local Group dwarf galaxies can be divided into two main classes: gas-rich systems (called ``dwarf irregulars'' when
forming stars, due to the irregular optical appearance caused by patches of recent star formation) and gas-deficient,
passively evolving ones (``classical dwarf spheroidals'' or ``dwarf ellipticals'', at the bright end; ``ultra-faint'' at the low 
luminosity end).
While the former are typically found in isolation, the latter are
in general satellites of the large Local Group spirals.

The fact that Local Group gas-rich and passive dwarfs share similar 
scaling relations \citep[see e.g.][]{mateo1998, tolstoy+2009}, 
appear to follow the same stellar mass-metallicity relation 
\citep{kirby+2013} - but inhabit clearly different environments -
already hints to similar formation mechanisms, albeit with environmental effects being 
relevant for the evolution of these galaxies.

To what extent and in what way dwarf galaxies are shaped by environment, it is still matter of debate. Theoretical studies 
show that strong tidal interactions with the host galaxy, coupled to ram-pressure stripping, could even turn a 
rotationally supported, gas-rich disky dwarf galaxy into a non-rotating, gas-deficient spheroidal system  
\citep[e.g.][]{mayer+2006}. If such ``tidal stirring'' is what produced the observed morphology-density relation, then  
the present-day structural, kinematic, and dark matter properties of passive dwarfs might be considerably different 
than prior to their interaction with the host. 
On the other hand, on the basis of lifetime star formation histories from very deep 
colour-magnitude-diagrams, \citet{gallart+2015} propose that 
the environment {\it at birth} is what sets dwarf galaxies on a given evolutionary path.

%Spectroscopic studies of the evolved stellar component of these systems 
%Even though currently available spectroscopic data-sets of individual stars in dwarf galaxies vary wildly in 
%terms of sample sizes and spatial coverages between nearby systems (MW satellites), and 
%more distant ones (M31 satellites and the isolated systems), 

Detailed studies of the kinematic and metallicity properties of the various types of 
dwarf galaxies inhabiting high and low density environments are clearly crucial for determining what 
properties are a common characteristic of this galaxy population and which ones are externally induced. 
To date, the large scale kinematic and metallicity properties of
the majority of Milky Way (MW) ``classical'' dwarf spheroidal galaxies (dSphs) have been studied in detail thanks to wide-area studies
based on samples of hundreds of individual stars 
\citep[e.g. to mention a few][]{kleyna+2002,tolstoy+2004,wilkinson+2004,munoz+2005,coleman+2005,battaglia+2006,koch+2006,walker+2006,faria+2007,koch+2007,koch+2008,battaglia+2008b,walker+2009, battaglia+2011,hendricks+2014,hendricks+2014b,walker+2015}.
The star formation and chemical enrichment history of MW dSphs vary spatially within these small galaxies
\citep[see also][]{deboer+2012a, deboer+2012b},
so that it is clear that MW dSphs cannot be defined by one number for their age or metallicity as was initially thought.
It is still to be determined though whether the observed age and metallicity gradients are caused by internal or external effects. 
%The presence of multiple ``chemo-dynamical'' stellar components has been uncovered in some of these objects, with metal rich stars
%generally more centrally concentrated and kinematically colder than metal poor stars
%\citep{tolstoy+2004,battaglia+2006,battaglia+2011,walker+penarrubia2011,amorisco+evans2012}. 

Gathering similarly large, spatially extended samples for the isolated Local Group dwarfs, as well as
M31 satellites, is more challenging and time consuming due to their larger heliocentric distances.
Nevertheless, notable efforts in this direction have recently taken place \citep[e.g.][]{ho+2012, collins+2013, kirby+2013, kirby+2014}.
In particular, both the kinematics and metallicity properties of WLM, a dwarf irregular (dIrr) galaxy at the edge of the Local Group, has been extensively
studied from a sample of $\sim$200 red giant branch (RGB) stars. This dIrr has a shallower metallicity gradient when compared with most well studied MW dSphs, but similar to those of larger and rotation supported systems as the Large and Small Magellanic Clouds \citep{leaman+2013}, 
suggesting that mass and angular momentum might play a role in driving the presence/absence of metallicity gradients \citep{schroyen+2011}. 

While it had previously been thought that dSphs were exclusively pressure-supported, hints of rotation were found in the outer parts of the MW dSph Sculptor \citep{battaglia+2008}, as well as in isolated dSphs (Cetus: \citealt{lewis+2007}; Tucana: \citealt{fraternali+2009}). 
Overall though, the stellar component of most Local Group dwarf galaxies studied so far is found to be dispersion-supported, 
possibly independently on morphological type and environment \citep{wheeler+2015}.
%\footnote{We note that the analysis of Wheeler et al. suggests a lower rotational-vs-dispersion support for Cetus, Tucana and WLM, with respect to the original studies \citep{lewis+2007, fraternali+2009, leaman+2009}.}
Even though sample sizes and spatial coverages of current spectroscopic samples of individual stars in Local Group dwarfs are rather inhomogeneous,
this result suggests that most dwarf galaxies 
form as thick, dispersion-supported systems and that, if passively evolving dwarfs originate from gas-rich systems, such transformation may occur just by removal of the gaseous component. 
Interestingly, so far only one Local Group dwarf galaxy, the And~II dSph, has been found to display prolate rotation \citep{ho+2012}, attributed to a merger with another small galactic system \citep{amorisco+2014, lokas+2014}.

In this article, we focus on the Phoenix (Phx) dwarf galaxy, which belongs to the 
handful of Local Group dwarf galaxies displaying intermediate properties between dIrrs and dSphs 
(and therefore called ``transition type dwarfs'', dIrr/dSph).
These are particularly interesting objects since they are
thought to be in the process of losing their gas, and therefore about to evolve into passive systems.
Being directly comparable to classical dSphs in terms of luminosities and sizes but typically 
found in less dense environments, dIrr/dSph offer the possibility of looking at the physical characteristics of
dSphs-like systems in absence of strong interactions with large galaxies. 

%In this article we extend this successful approach applied to dSphs to study the transition type Phoenix (Phx) dwarf galaxy.
Located at $\sim450$\,kpc from the Sun \citep{vanderydt+1991}, Phx is the closest dIrr/dSph and therefore a 
good candidate for accurate studies. It is found outside the virial radius of the MW and M\,31, 
hence arguably environmental effects were not the main driver in its evolution.
Phx has a luminosity similar to the Sculptor dSph ($\rm{L_V} = 0.9 \times 10^6\,L_{\odot}$ and $2.15\times10^6\,\rm{L_{\odot}}$, respectively; \citealt{mateo1998}) and, like Sculptor, it formed most of its stars more than $10$\,Gyr ago \citep{hidalgo+2009}.
However, unlike Sculptor, Phx has been forming stars recently, up until 100 Myr ago \citep{young+2007}.
%Also, HI clouds are likely to be associated to the system \citep{germain+1999}.
This shows that Phx was typically able to retain some gas, probably because of its larger distance from the MW and/or a larger total mass.

To date, spectroscopic studies of resolved stars in Phx have been scarse. Even its optical systemic velocity 
is uncertain, with conflicting results from two spectroscopic studies based on only a few RGB stars in the central region of Phx
 ($-52\pm6$\kms\,, \citealt{gallart+2001}; $-13\pm9$\kms\,, \citealt{irwin+tolstoy2002}). This makes the physical association with a nearby 
HI cloud \citet{young+lo1997, germain+1999} still debatable.  
The available photometry shows that the young stars are centrally concentrated, and hints at differences in the spatial distribution of the older and intermediate age stars,
as observed for several dSphs \citep{battaglia+2012, hidalgo+2013}. In addition, the 
spatial distribution of the young stars is tilted by $\sim90^{\circ}$ with respect to Phx main body and aligned with the projected 
minor axis \citep{martinez-delgado+1999, hidalgo+2009, battaglia+2012}. Interestingly, signs of kinematic properties varying with position angle
have been reported in a conference proceeding by \citet{zaggia+2011}. Nothing is known about the metallicity properties of this galaxy nor whether it is
rotationally or pressure-supported. 

\begin{table}
	\centering
	\caption{Adopted parameters for Phx.}
	\label{tab:parameters}
	\begin{tabular}{ccc} % four columns, alignment for each
	\hline
Parameter & Value &  Ref. \\
        \hline
$\alpha_{\rm J2000}$ & $01^h\,51^m\,06^s$ &  \citet{mateo1998} \\
$\delta_{\rm J2000}$ & $-44^{\circ}\,26'\,41''$ &  \citet{mateo1998} \\
ellipticity\tablefootmark{a} & 0.3 &  \citet{battaglia+2012} \\
P.A. & $5^{\circ}$ &  \citet{battaglia+2012} \\
$R_{core}$ & $1.79'$ & \citet{battaglia+2012} \\
$R_{tidal}$ & $10.56'$ &  \citet{battaglia+2012} \\
$R_{1/2}$ & $2.30'$ &  \citet{battaglia+2012} \\
$R_{S}$ & $1.67'$ & \citet{battaglia+2012}\tablefootmark{b} \\
$m$   & $0.88$ &  \citet{battaglia+2012}\tablefootmark{b} \\
$V_{HB}$ & $23.9$ & \citet{holtzman+2000} \\ 
dm$_0$ & 23.09$\pm$0.10 & \citet{mcconnachie2012} \\
E(B-V) & 0.016 &  \citet{mcconnachie2012}\\
	\hline
	\end{tabular}
\tablefoot{
\tablefoottext{a}{$e=1-\frac{b}{a}$}
\tablefoottext{b}{We re-derived the Sersic profile parameters (characteristic radius $R_S$ and power $m$) since we noticed that some duplicates were present in the 2012 photometry in the overlapping regions between the different pointings. Removing these duplicates from the photometric catalogue does not lead to any significant changes in terms of structural properties.} 
}
\end{table}

Here we extend to Phoenix the successful approach applied to MW dSphs, by deriving its large scale metallicity and kinematic properties
from a wide-area spectroscopic survey of $\sim$200 red giant branch stars in the Ca~II triplet (CaT) region. In Sects.~\ref{sec:obs} and \ref{sec:ew} we present the
data-set, the data reduction procedure and the determination of metallicities ([Fe/H]) from the nIR CaT lines. In Sect.~\ref{sec:mem} we discuss the
membership selection. Sects.~\ref{sec:kinematics} and \ref{sec:metallicity} contain the analysis of the internal kinematic and metallicity properties.
Finally, we conclude and summarise the main results in Sect. \ref{sec:conclusions}. The Phx parameters adopted throughout this work are summarised in Table~\ref{tab:parameters}.

%{\it The document doesn't compile without this text.}
%{\it The document doesn't compile without this text.}
%{\it The document doesn't compile without this text.}

\section{Observations and data reduction} \label{sec:obs}

\begin{figure}
	\includegraphics[width=8.1cm]{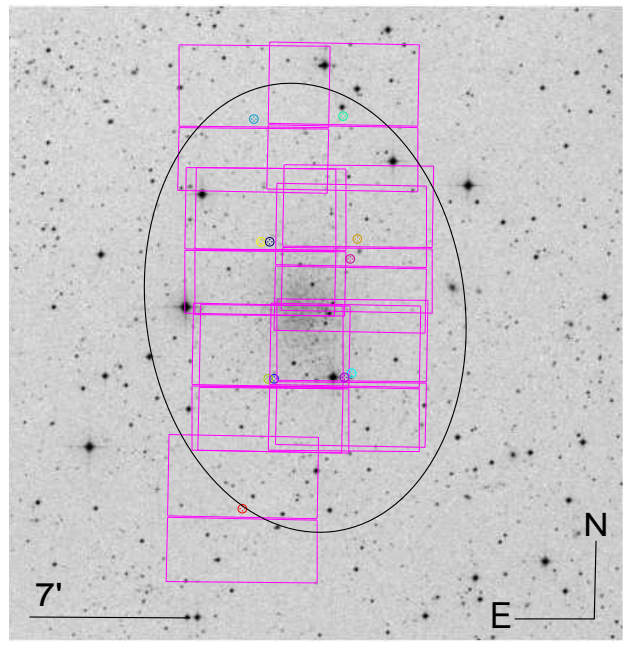}
        \caption{DSS image of the Phx dwarf galaxy. The field of view of FORS2 in MXU mode
          is overplotted for the eleven pointings from the P83 observing run. The centres of the different pointings are marked with different colour symbols. The ellipse shows the nominal tidal radius, position angle and ellipticity of Phx \citep{battaglia+2012}.}
    \label{fig:Phx-FoV}
\end{figure}

\begin{figure}
	\includegraphics[width=\columnwidth]{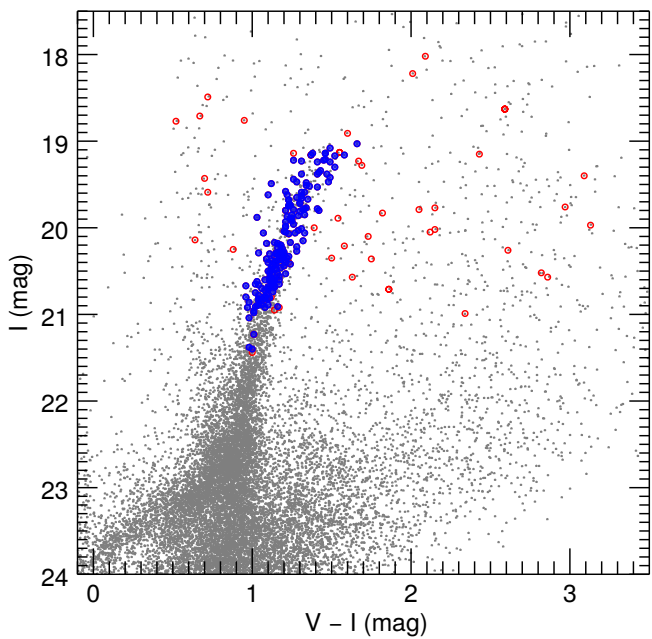}
    \caption{Colour-magnitude diagram of the Phx dwarf galaxy from VLT/FORS2 photometry \citep{battaglia+2012}. The FORS2 MXU targets are marked with open, red circles and the probable Phx member stars are shown with solid, blue symbols.}
    \label{fig:Phx-CMD}
\end{figure}

Observations have been taken with the FORS2 instrument in multi-object spectroscopic mode 
with exchangeable masks. FORS2 is a focal reducer and visual-light multi-purpose instrument that offers imaging, polarimetry, low-dispersion long slit and multi-object spectroscopy \citep{appenzeller+1998} on the VLT at the ESO Paranal Observatory. 

The bulk of our analysis is based on a data set taken in service mode between 18--29 September 2009 within ESO Programme ID 083.B-0252(B) (PI: Battaglia), when FORS2 was mounted at the Cassegrain focus of Antu (UT1). Within this programme we obtained spectra of $254$ stars with I\,$<21.5$\,mag distributed over 11 different pointings, covering a total area of $\sim 10\arcmin \times 20\arcmin$
across the galaxy (Fig.~\ref{fig:Phx-FoV}). 11 stars were targeted more than once in different masks and could thus be used to assess the measurement errors from independent observations.

This main data set is complemented with another one taken in visitor mode on 23 August 2003 within ESO Programme ID 71.B-0516 (PI: Cole), when FORS2 was mounted at the Cassegrain focus of Yepun (UT4). We observed one position in central Phoenix targeting 40 stars within 1 magnitude of the tip of the RGB (19.2 $\lesssim$ I $\lesssim$ 20). There are 14 stars in common between the two data sets.
Both data sets are analysed and presented here for the first time.
%The seeing varied from 0.8--1$\arcsec$, so that radial velocity corrections due to slit-centering errors were required. 
 
The observing setup was identical during both runs: we used the 1028z grism in combination with the OG590+32 order separation filter (wavelength coverage 7730 - 9480\,\AA, dispersion $0.42$\,\AA\,px$^{-1}$, spectral resolution $R = \lambda/\Delta \lambda = 2560$) and the standard resolution mode of the red-sensitive MIT detector (two $2\rm{k}\times4\rm{k}$\,px CCDs mosaic, with 2x2 binning that yields $0.25\arcsec\,\rm{px}^{-1}$ spatial scale). We used the Mask eXchange Unit (MXU) mode to design  $1\arcsec$ wide ($10\arcsec$ long) slits across the $6.8\arcmin \times 6.8\arcmin$ field of view. Targets were selected based on FORS2 pre-imaging taken in V and I-band prior to each spectroscopic run. The pre-imaging observations used for the selection of the spectroscopic targets for the programme ID 083.B-0252 (Period 83 or P83) are described in \citet{battaglia+2012}. The priority was given to targets identified along the Phx RGB (Fig.~\ref{fig:Phx-CMD}), with few stars added outside this colour range, when the slits could not be allocated to a suitable Phx RGB candidate. 

In addition to Phoenix stars, observations were also taken with MXU masks designed to target RGB stars in Galactic globular clusters in order to calibrate the relationship between Ca~II triplet (CaT) equivalent width (EW) and metallicity \citep{leaman+2009, swan+2016}.
We observed two clusters that bracket the expected metallicity range of Phx: 47\,Tuc, [Fe/H]\,$=-0.72$\,dex and M\,15, [Fe/H]\,$=-2.37$\,dex (\citealt[][2010 edition]{harris1996}\footnote{\url{http://www.physics.mcmaster.ca/resources/globular.html}}), each with one MXU mask per cluster. 

\begin{table}
	\centering
	\caption{Observing log.}
	\label{tab:observing-log}
	\begin{tabular}{ccccc} % four columns, alignment for each
		\hline
		Field\tablefootmark{1}  & RA2000 & DEC2000 & Date & Exp. time \\
		      &  [deg] & [deg]   & Sep. 2009 & [s]             \\
		\hline
Phx-14     &   27.84209	&   $-$44.60355  & 27 \& 28 & 2$\times$2730 \\
Phx-10-1   &   27.80934	&   $-$44.50062  & 21 \& 25 & 3$\times$2730 \\
Phx-10-2   &   27.81683	&   $-$44.50057  & 23          & 1$\times$2730 \\
Phx-11-1   &   27.72505	&   $-$44.49538  & 26          & 3$\times$2730 \\
Phx-11-2   &   27.73237	&   $-$44.49931  & 18          & 2$\times$2730 \\
Phx-06-1   &   27.82667	&   $-$44.39353  & 26          & 3$\times$2730 \\
Phx-06-2   &   27.81574	&   $-$44.39362  & 22          & 2$\times$2730 \\
Phx-07-1   &   27.72061	&   $-$44.39015  & 24 \& 25 & 4$\times$2730 \\
Phx-07-2   &   27.72773	&   $-$44.40565  & 27 \& 28 & 2$\times$2730 \\
Phx-02     &   27.83579	&   $-$44.29739  & 27          & 1$\times$2730 \\
Phx-03     &   27.73828	&   $-$44.29482  & 29          & 1$\times$2730 \\
47\,Tuc    &    5.65952   &   $-$72.07518  & 23          & 3$\times$100 \\
M\,15      &  322.53856   &   $+$12.17369  & 24          & 3$\times$300 \\
\hline
          &               &               & Aug. 2003 &               \\
Phx (centre) & 27.76975 & $-$44.42647 & 23 & 6$\times$2000 \\
\hline
	\end{tabular}
\tablefoot{
\tablefoottext{1}{Fields are ordered from South to North and East to West.}
}
\end{table}

The observing logs including number of exposures and exposure times for each pointing and the central pointing coordinates for 2009 (referred to as P83 according to ESO parlance) and 2003 (referred to as P71 run below) observing runs are in Table~\ref{tab:observing-log}.  

In the following subsections we describe our data reduction procedure and radial velocity measurements, emphasising in particular (small) differences between the two data sets and the strategy adopted to produce the final homogeneous sample used for the kinematic and metallicity analysis. 

\subsection{Data reduction}

Data reduction and extraction of the spectra was performed within IRAF\footnote{IRAF is distributed by the National Optical Astronomy Observatories, which are operated by the Association of Universities for Research in Astronomy, Inc., under cooperative agreement with the National Science Foundation. \url{http://iraf.noao.edu/}}.
Using standard IRAF procedures,  master bias and bias-subtracted, normalised master flat-field images, were created from standard calibrations (5 frames each for bias and lamp flats) taken typically in the morning following the science observations. These master calibrations were used to correct the two-dimensional (2D) science and wavelength calibration exposures. We then applied the spectroscopic modification of the {\it L.\,A.\,Cosmic}\footnote{\url{http://www.astro.yale.edu/dokkum/lacosmic/}} algorithm \citep{vanDokkum2001} to remove cosmic rays from the science frames.

At this point only the individual exposures of each of the two calibration globular clusters masks were median combined. Due to the short interval of time in which these exposures were recorded, there were practically no shifts nor tilts among them. This was, however, not the case for the Phx exposures, where some shifts and PSF variations due to changing seeing over the night or between different nights were evident.

We used custom designed IRAF scripts to trace the apertures and correct for distortions the science frames and associated arc-lamp frames. Data taken between 18--28 September 2009 showed strongly tilted spectra both in calibration (arc-lamp) and science frames. The tilt was of the order of $\sim 50$ binned pixels between start and end of the CCD, and was caused by grism misalignment. To correct for this we applied the wavelength solution directly on the rectified 2D spectra before extraction. More specifically, we cut the individual rectified 2D spectra from the science and arc-lamp images, identified the lines from the arc-lamp with IRAF {\it identify} and {\it re-identify} tasks and then traced their position along the slit to obtain a 2D model of the wavelength calibration for each individual slit with {\it fitscoord} and {\it transform} tasks. 
Applying this model we obtained wavelength calibrated 2D spectra with straightened sky lines, based on which subsequent sky-subtraction could be substantially improved.
We continued with optimal extraction and sky subtraction using the standard IRAF procedure {\it apall}, which results in the optimally extracted 1D science spectrum accompanied with additional extensions that have the extracted sky spectrum and the error spectrum -- the flux uncertainty at every pixel. At this point we performed an additional refinement to the wavelength calibration using the measured positions of the sky lines in the extracted sky spectra, such that radial velocities measured from the sky spectra had a mean of $0\pm2$\kms. The resulting 1D spectra were then continuum normalised using a high order polynomial fits within the IRAF task {\it continuum}. 

We note that for the P71 single MXU mask data reduction was slightly different: the extracted sky spectra were used to determine the dispersion solution of the spectra individually, and then the spectra were continuum normalised, shifted to a common wavelength solution and combined to produce the final spectrum.

\subsection{Radial Velocity measurements and wavelength calibration accuracy}

The good centring of the stars in the slits in low resolution spectroscopy is crucial for the accurate determination of the radial velocities, in particular when seeing is of the order of or smaller than the slit width (which was the case for most of our observations). We performed a consistency check for the centring of each star's centroid on the slit using the through-slit images taken before the spectroscopic exposures. This showed that most masks were well centred to within $0.1\pm0.1$\,px (median and standard deviation; $1\,\rm{px} = 28$\kms).
Only in two exposures of one P83 mask we measured a significant slit centre offset of $0.3\pm0.1$\,px, as well as in the two globular cluster masks ($0.7\pm0.2$\,px and $0.2\pm0.2$\,px for 47\,Tuc and M\,15, respectively). Likely the worse slit alignment of globular cluster masks was due to bright targets, that appeared to fill the entire slit even with very short exposure, unless one uses additional filter or adjusts and carefully examines the cuts in the through slit images. These exposures were corrected for the corresponding radial velocity offsets.

We measured the radial velocities from the 1D wavelength calibrated spectra of each individual exposure using the IRAF {\it fxcor} task, which uses a Fourier cross-correlation to determine the Doppler offset of the spectrum of interest compared to a template spectrum. As a template, we used a synthetic spectrum derived from an interpolated Kurucz stellar atmospheric model and convolved to match the FORS2 spectral resolution with similar parameters as expected for the Phx targets (i.e. RGB stars of moderately low metallicity - ${\rm logg}=1.0$, ${\rm T_{eff}} = 4000$\,K, [Fe/H]$=-1.5$\,dex). Then we calculated the heliocentric radial velocities using the {\it rvcorrect} task in IRAF.
The {\it fxcor} task also computes velocity uncertainties based on the fitted correlation peak height and the antisymmetric noise \citep[see][]{tonry+davis1979}, which turned to have a median value of $\pm12$\kms.
To this we added in quadrature the velocity uncertainty due to the wavelength calibration (median value of $\pm2$\kms) and the slit centring  ($\pm4$\kms) systematic errors. 
The final velocity of each star, and the corresponding velocity uncertainty, was calculated as the weighted mean and error of the weighted mean of the velocities from the individual exposures. As discussed below we have a better handle on the velocity
uncertainties from individual spectra, rather than measuring radial velocities from the stacked spectra.

Finally, the spectra of the individual exposures were Doppler corrected and combined using a weighted average where more than one exposure per star was available. Due to the small number of individual exposures per star (typically 2 or 3; see Table \ref{tab:observing-log}) a median combination was not justified. In order to obtain an error spectrum for each reduced science spectrum, the individual $\sigma$-spectra were divided by the polynomial used for the continuum normalisation of the science spectra and combined using the formula for estimating the error of a weighted mean ($\sigma^2(\lambda) = 1 / \Sigma[1/\sigma_i^2(\lambda)]$).
The median signal-to-noise ratio (SNR) of our spectra is $23$ per px for the Phx masks and $77$ and $90$ per px for 47\,Tuc and M\,15, respectively. We note that we use the regions between $8575$ and $8630$\,\AA~and $8700$ and $8750$\,\AA~to track the continuum level and uniformly estimate the SNR for each spectrum.

\begin{figure}
	\includegraphics[width=\columnwidth]{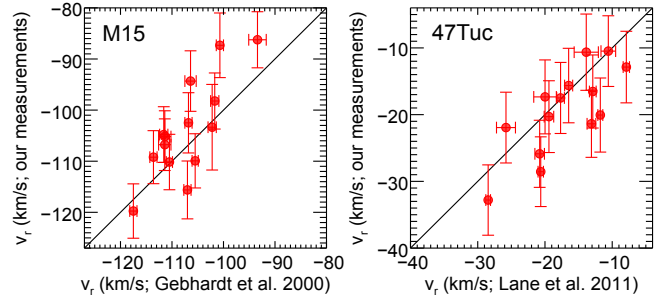}
    \caption{A comparison of our radial velocity measurements for the two globular clusters with high resolution measurements for the same stars from \citet[][M\,15]{gebhardt+2000} and \citet[][47\,Tuc]{lane+2011}. The black line is not a fit, but a 1:1 relation.}
    \label{fig:RV-comparison}
\end{figure}

\begin{figure}
	\includegraphics[width=\columnwidth]{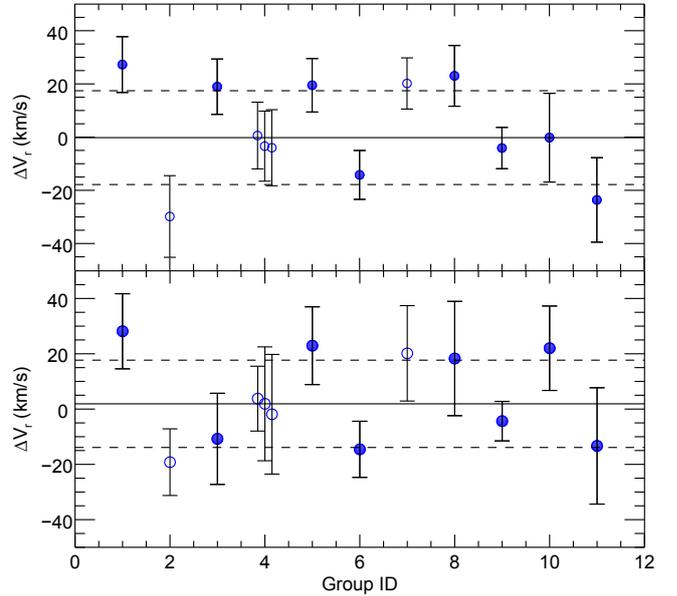}
    \caption{The difference in the radial velocity measurements for the 11 stars that were targeted in more than one mask in P83. The star in Group 4 was observed in three different masks and all difference combinations are plotted. Phx probable members are plotted with solid blue symbols, while non-members are plotted with open symbols. The error bars denote the quadrature combined uncertainties of both measurements.
    The median of the distribution is indicated with a solid black line and the $\pm1\sigma$ level is indicated with dashed lines.
    {\it Upper panel:} Velocities are measured after stacking of the individual exposures.
    {\it Bottom panel:} Velocities are measured from individual exposures and then the weighted mean values and their uncertainties are calculated.}
    \label{fig:common-stars}
\end{figure}

\begin{figure}
	\includegraphics[width=\columnwidth]{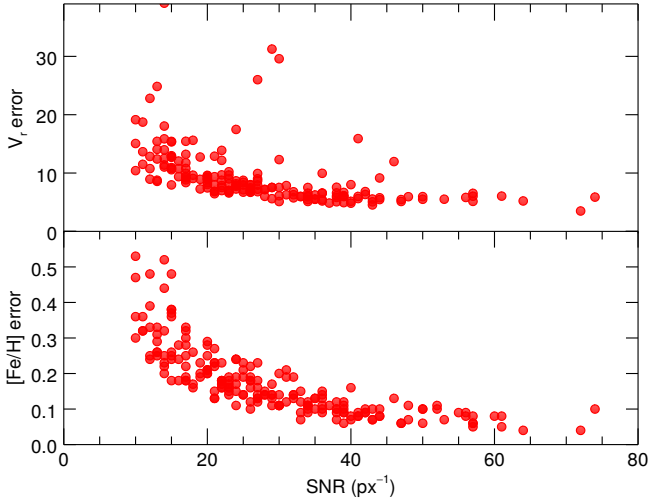}
    \caption{The radial velocity and metallicity uncertainties of the Phx member stars as a function of the SNR in the P83 data set.}
    \label{fig:snr-v-feh}
\end{figure}

We matched the observed stars in the two globular clusters with high resolution radial velocity catalogues  \citep[][for M\,15 and 47\,Tuc, respectively]{gebhardt+2000,lane+2011} to assess the accuracy of our measurements. The results of the comparison are presented in Fig. \ref{fig:RV-comparison}. There is no significant systematic offset between our radial velocity estimates and the results from high resolution studies.
The mean radial velocities from our FORS2 observations for M\,15 and 47\,Tuc are $-104.0\pm2.5$\kms~and $-19.6\pm1.7$\kms, respectively, in good agreement with the corresponding values published in the \citet[][2010 version]{harris1996} catalogue
($-107.0\pm0.2$\kms~for M\,15 and $-18.0\pm0.1$\kms~for 47\,Tuc).
We note that the globular clusters spectra have significantly higher SNR on average than our Phx spectra and hence also smaller velocity uncertainties.

For the Phx targets we tested the accuracy of the velocity measurements using 11 stars that were observed in more than one mask. 
We compared the difference between the derived velocities from each mask for these stars in common (Fig. \ref{fig:common-stars}) both from measurements performed on stacked spectra and on individual exposures. The estimated uncertainties in the first case (measurement from stacked spectra) are too small compared to the standard deviation of the distribution of $17$\kms, while the median, quadrature combined error for the presented differences is $11$\kms. On the other hand, the second case (RV measurements from individual exposures) results are in much better agreement, with the standard deviation of the distribution amounting to $15$\kms and the median, quadrature combined error for the presented differences being $15$\kms too. This corroborates our choice of adopting as final velocities those
derived as weighted mean of the measurements from individual exposures, rather than from the stacked spectra.

In the upper panel of Fig. \ref{fig:snr-v-feh} we show the final adopted uncertainties of the radial velocities for all probable Phx member stars as function of the SNR of the spectra; the  median velocity uncertainty is $8$\kms. 
The detailed kinematics of the Phx dwarf is presented in Section~\ref{sec:kinematics}.

\subsection{Comparison of the P71 and P83 data sets}

The radial velocity and CaT EW measurements were performed in the same way for the two data sets. Radial velocities were determined by cross-correlation with the IRAF task {\it fxcor} adopting the same synthetic template spectrum and the slit-centering corrections were derived by comparing the centroid of starlight to the centroid of the sky light seen through the slit mask with the grating removed. The corrections for the P71 single MXU setup were between $0.1$ -- $0.4$ pixel after taking into account the fact that the telescope was moved blindly by $0.5$\,px after the last through slit image in order to compensate a larger initial offset. The resulting systematic negative offset averages at $-$8.6\kms with a star-to-star dispersion of $\pm2.6$~\kms ($\pm$0.1~px). The determination of the EWs of the CaT lines is described in Sect.~\ref{sec:ew}.

\begin{figure}
	\includegraphics[width=\columnwidth]{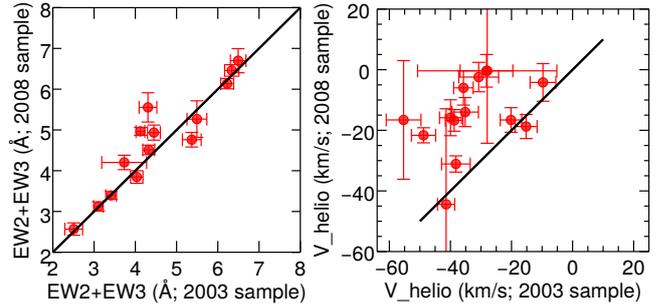}
    \caption{CaT Equivalent widths ({\it left panel}) and radial velocities ({\it right panel}) comparison for the 14 stars in common in the P83 and P71 spectroscopic sets. Black lines indicate equality.}
    \label{fig:cole-comp}
\end{figure}

Fig. \ref{fig:cole-comp} shows the comparison between the velocities and CaT EWs for
14 stars in common in the P71 and P83 spectroscopic samples. While there is a generally good agreement
between the measured CaT EWs, there is a clear discrepancy between the velocity measurements, which
appears mostly, but not entirely, due to a different velocity zero-point. FORS2 is a low-resolution spectrograph, not optimised for accurate velocity measurements, suffering from flexures. Besides it was on two different telescopes at the time of the two runs. We could not identify exactly the origin
of the velocity difference but it most likely is due to the fore-mentioned blind adjustment of the telescope pointing after the last through-slit image. Thus, we prefer to rely only on the larger and homogeneous P83 sample for the kinematic analysis. On contrary, we deem it safe to combine the two samples for the determination of the CaT EWs, because we measure CaT EWs on the Doppler corrected spectra.
For the stars in common between the P71 and P83 data sets, we adopted the weighted mean values of their CaT EWs.

\section{Ca~II triplet equivalent width and metallicity calibration}
\label{sec:ew}

The CaT is the most prominent spectral feature in the near-IR spectra of cool stars.
The three lines are centred at $8498.02$\,\AA, $8542.09$\,\AA, and $8662.14$\,\AA\, and their strengths have been extensively used as reliable indicators of the metallicity ([Fe/H]) of individual RGB stars in a large variety of stellar systems: old and relatively metal poor globular clusters \citep[e.g.][]{armandroff+zinn1988,rutledge+1997}, more metal rich open star clusters \citep[e.g.][]{cole+2004,warren+cole2009,carrera2012}, but also dwarf galaxies with complex star formation histories and extended metallicity and age spreads \citep[e.g.][]{armandroff+dacosta1991,tolstoy+2001,cole+2005,koch+2006,carrera+2008b,carrera+2008a,battaglia+2008,hendricks+2014}. The reliability of the method has been tested via comparison of the CaT [Fe/H] with direct [Fe/H]
measurements 
from high resolution spectra of the same stars
in systems with composite stellar populations exhibiting a range of metallicities \citep[e.g. dwarf galaxies, the MW bulge][]{battaglia+2008, vasquez+2015}.

A review of the variety of CaT EW-[Fe/H] calibrations introduced
in the literature is beyond the scope of this work and we refer the reader to the various studies
for details on the method. Here we use our globular cluster observations to briefly test the
\citet{starkenburg+2010} and \citet{carrera+2013} relations, which are calibrated
over a large range in [Fe/H] (-4 $<$ [Fe/H] $< -0.5$ and -4 $<$ [Fe/H] $< +0.5$, respectively) that is suitable for our target, and take into account
the fact that the linearity of the calibration breaks down at [Fe/H]\,$\lesssim-2$\,dex.

Both calibrations follow the same functional form:
\begin{equation}
 {\rm[Fe/H]} = a + b(V-V_{\rm{HB}}) + c\Sigma_{\rm{EW}} + d\Sigma_{\rm{EW}}^{-1.5} + e\Sigma_{\rm{EW}}(V-V_{\rm{HB}}),
\end{equation}
but assume different coefficients ($a,b,c,d,e$). The use of the magnitude difference with respect to the horizontal branch (V-V$_{\rm{HB}}$) was 
introduced by \citet{armandroff+dacosta1991} in order 
to compare RGB stars of different luminosities, independently on distance and reddening. A disadvantage,
especially in complex stellar populations, is
the difficulty to pinpoint a unique HB magnitude for the
entire system; however this effect is expected to introduce an error of $\sim0.05$\,dex, much smaller than the
uncertainty intrinsic to the method itself \citep[e.g.][and references therein]{cole+2004}.
We adopted $V_{\rm{HB}} = 23.9$\,mag \citep{holtzman+2000} for the Phx dwarf. $\Sigma_{\rm{EW}}$ is a 
linear combination of the EW of the CaT lines: both the \citet{carrera+2013} and the \citet{starkenburg+2010}
calibration use an unweighted sum of the
EWs of the CaT lines, but while the former uses the three lines, the latter uses only the two strongest lines
due to concerns that the weakest line at $8498$\,\AA~might introduce more scatter to the calibration
\citep[see also][]{koch+2006,battaglia+2008,hendricks+2014}. It should be noted that
also the systemic velocity of the system under consideration
may play a role in the choice of the lines to use, as e.g. one of the lines may be more affected by OH-band sky residuals.
Finally, in order to measure the EW of the individual CaT lines one has the choice between a direct flux integration
or fitting a functional form to the line shape to be integrated over a band-pass. 
In the following we will adopt the prescriptions from \citet{starkenburg+2010} and \citet{carrera+2013}  in terms
of the band pass over which to integrate the line flux, the linear combination of CaT EWs to be used, and obviously the
($a,b,c,d,e$) coefficients, while experimenting with the method of flux integration (fit or direct flux integration).

We developed our own algorithm to measure the EWs of the CaT lines from the
continuum normalised stacked spectra
using the {\it mpfit}\footnote{\url{http://purl.com/net/mpfit}} package in IDL \citep{markwardt2009}.
We tested EW determination by fitting the CaT lines with pure Gaussians and with 
true Voigt profiles (a convolution between a Gaussian and a Lorentzian profile), and adopted   
the $\sigma$-spectra as the flux uncertainty at each wavelength pixel.
Additionally, we calculated a simple pixel-to-pixel integration of the line flux.
We note that we trusted the overall continuum normalisation of our spectra, which we deem more robust compared to using narrow custom defined continuum windows in the calculations, as is also done in e.g. \citet[][]{leaman+2009} with similar data sets.

\begin{figure*}
	\includegraphics[width=18cm]{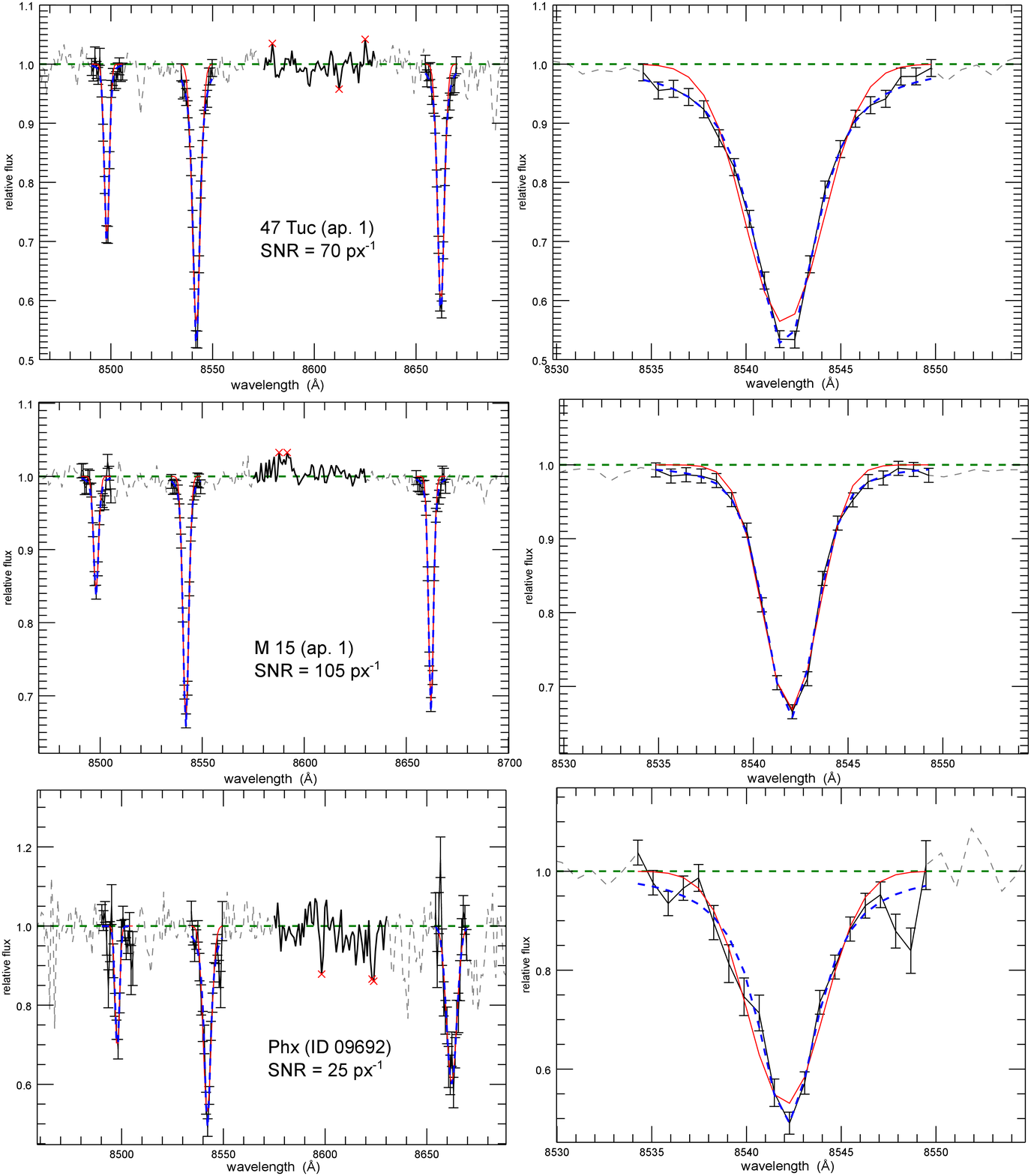}
    \caption{Example spectra of one 47\,Tuc (upper row), M\,15 (middle row), and Phx (bottom row) star. The left column shows the three Ca lines and the right column shows only the middle $8542$\,\AA~line. The best fit Gaussian profile is indicated with a solid red line and the Voigt profile is indicated with a dashed blue line. The observed spectrum is marked with a dashed gray line. A solid black line with error bars indicates the line band pass region used for the profile fitting and flux integration. One of the defined continuum regions for SNR estimation is marked with a solid black line, and the continuum level with a green dashed line. Red crosses indicate continuum points 
rejected by a $\sigma$-clipping algorithm.}
    \label{fig:cat-fits}
\end{figure*}

In Fig. \ref{fig:cat-fits} we show example spectra and the fitted model functions for one representative star
from the two globular clusters and one star from the Phx galaxy with a typical SNR of $25$ per pix. 
Table~\ref{tab:calibrations} summarises the comparison between the CaT [Fe/H] that we obtain for our calibrating
globular clusters 47\,Tuc and M\,15 when using the two CaT EW-[Fe/H] calibrations for the various  EW measurement
methods (pixel-to-pixel integration, Gaussian and Voigt fits). 
%Owing to the relatively narrow band widths defined in the \citet{starkenburg+2010} calibration, the difference is substantial and the EWs are rather underestimated. Using a CaT-metallicity calibration based on...
In line with other studies in the literature, it can be clearly seen that the Gaussian fit is a poor
representation of the line shape at high metallicity, missing the wings of the line,
which then leads to underestimating the [Fe/H] abundance of 47\,Tuc for both CaT 
calibrations; on the other hand, the Voigt fits not only offer a suitable representation of the CaT line profiles, but are
also less prone to subtle fluctuations of the flux due to the presence of weak lines and noise that affects more the
direct flux integration; hence we adopt the EWs inferred
from Voigt fits for the rest of the discussion in this paper.

The \citet{carrera+2013} calibration is based on observations from multiple instruments including FORS2 but with higher spectral resolution of $\sim5000$. They adopted the line band windows of \citet{cenarro+2001}, which are broad ($29$\,\AA, $40$\,\AA, $40$\,\AA~for the three lines, respectively) and account for the extended line wings at high metallicity.
In our analysis such broad band passes turn out to be, however, 
very sensitive to small uncertainties of the continuum zero point. 
Our Voigt profile fit procedure predicts artificially large Lorentzian broadening for the metal rich stars of 47\,Tuc, which leads to very large EWs and hence higher than expected metallicity. 
At the metal poor regime, the calibration reproduces well the metallicity of M\,15 (see Table \ref{tab:calibrations}).
We notice that we can reproduce the expected metallicities of the two globular clusters with the \citet{carrera+2013} calibration if we use the narrower line band passes proposed by \citet[][$16$\,\AA, $20$\,\AA, $18$\,\AA, respectively]{armandroff+zinn1988}.

The \citet{starkenburg+2010} relation, on the other hand, was not calibrated up to metallicities as high as in the \citet{carrera+2013} study and
uses band windows of $15$\,\AA~for the two stronger CaT lines following the approach described by \citet{battaglia+2008}.
The advantage of such narrow integration windows is that the calibration is not sensitive to small continuum uncertainties and effects such as random noise or weak spectral lines in the vicinity of the CaT.
It might, however, be less sensitive at high metallicities due to the exclusion of the extended Ca~II line wings from the EW integration.
Nevertheless, using this calibration we reproduced very well the metallicities of both 47\,Tuc at the high metallicity end and M\,15 in the low metallicity regime. Therefore we decided to adopt \citet{starkenburg+2010} calibration to derive the metallicities of the Phx stars, 
based on the EWs from the Voigt fits integrated over a band-pass of $15$\,\AA~.
The \citet{starkenburg+2010} calibration was obtained using synthetic spectra of resolution $R = 6500$. This is larger than our FORS2 resolution but we confirmed by examining synthetic spectra with the average parameters of the 47\,Tuc stars, convolved to different instrument resolutions, that the signal within the $15$\,\AA~band windows is practically the same.

We used two approaches to compute the uncertainties of the EW measurements for the observed stars.
The {\it mpfit} code provides formal $1\sigma$ errors based on the covariance matrix of the fitting parameters.
Since our line profile fits were weighted with the corresponding $\sigma$-spectra,
the errors calculated in this way should be representative of the true uncertainties.
As a second approach to compute the uncertainties, we used the empiric formula proposed by
\citet{cayrel1988} in the form from \citet{hendricks+2014}:
\begin{equation}
 \Delta_{\rm{EW}} = 1.725\,\rm{SNR}\,\sqrt{\sigma_{\rm{Gauss}}},
\end{equation}
where $\sigma_{\rm{Gauss}}$ is the $1\sigma$ broadening of the Gaussian profile fit,
which we assumed to be very close to the equivalent value of the Voigt profile fit.
The Cayrel formula is based solely on the resolution and the SNR of the spectra.
%A comparison between the two approaches is presented in Figure \ref{fig:ew-err}.
We find the uncertainties inferred from the two methods to be strongly
correlated but the errors based on the goodness of the fit are systematically larger
for both the good quality globular cluster spectra and the fainter Phx spectra.
We adopt the latter as the final uncertainties on the EW measurements.

The errors on the EWs of the two strongest Ca~II triplet lines were combined in quadrature and we used the bottom and upper EW boundaries to calculate the uncertainties on the metallicity estimates.
In the bottom panel of Fig. \ref{fig:snr-v-feh} we show the final [Fe/H] uncertainties for all probable Phx member stars, with a median
uncertainty of  $0.16$\,dex.

\begin{table}
	\centering
	\caption{Median metallicities ([Fe/H]) and measured spreads ($\sigma_{\rm{[Fe/H]}}$) for the two calibration globular clusters and the clean Phx sample according to different calibrations and EW measurement methods 
(``Sum'' refers to the direct pixel-to-pixel flux integration). }
	\label{tab:calibrations}
	\begin{tabular}{cccc} % four columns, alignment for each
		\hline
		Method  & 47\,Tuc & M\,15 & Phx\tablefootmark{2}  \\
		        & $-0.72$\tablefootmark{1} & $-2.36$\tablefootmark{1} & \\
		\hline
		\multicolumn{4}{c}{\citet{carrera+2013}\tablefootmark{3}} \\
		Gauss  & $-1.05\pm0.10$ & $-2.57\pm0.07$ &  \\
		Voigt  & $-0.52\pm0.13$ & $-2.41\pm0.09$ &  \\
		Sum    & $-0.42\pm0.18$ & $-2.42\pm0.16$ &  \\
		\multicolumn{4}{c}{\citet{carrera+2013}\tablefootmark{4}} \\
		Gauss  & $-1.05\pm0.10$ & $-2.57\pm0.07$ &  \\
		Voigt  & $-0.67\pm0.11$ & $-2.33\pm0.08$ &  \\
		Sum    & $-0.71\pm0.12$ & $-2.46\pm0.10$ &  \\
		\multicolumn{4}{c}{\citet{starkenburg+2010}} \\
		Gauss  & $-0.90\pm0.09$ & $-2.37\pm0.08$ & $-1.67\pm0.62$ \\
		Voigt  & $-0.71\pm0.09$ & $-2.27\pm0.10$ & $-1.53\pm0.59$ \\
		Sum    & $-0.68\pm0.09$ & $-2.26\pm0.11$ & $-1.50\pm0.62$ \\		
		\hline
	\end{tabular}
	\tablefoot{
	\tablefoottext{1}{[Fe/H] according to the catalogue of \citet[][2010 edition]{harris1996}.}
	\tablefoottext{2}{Only probable Phx member stars are included.}
	\tablefoottext{3}{Using the \citet{cenarro+2001} band windows as in \citet{carrera+2013}.}
	\tablefoottext{4}{Using the \citet{armandroff+zinn1988} band windows.}
	}
\end{table}

\begin{figure}
	\includegraphics[width=\columnwidth]{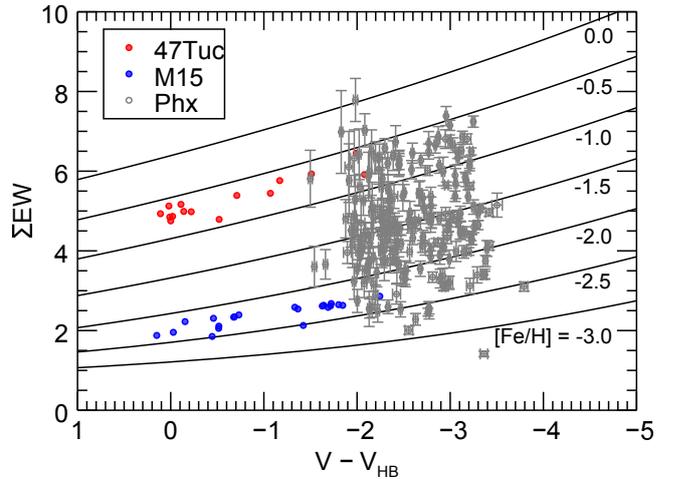}
    \caption{The sum of the EWs of the two stronger CaT lines as a function of $\rm{V-V_{HB}}$ for the clean Phx sample and the globular cluster stars. Lines of equal metallicity according to the \citet{starkenburg+2010} calibration are overplotted with solid black lines.}
    \label{fig:cat-calibration}
\end{figure}

In Fig. \ref{fig:cat-calibration} we plot $\Sigma_{\rm{EW}}$ as a function of the V magnitude above the HB. One can see the accurate metallicity estimates for the two calibration globular clusters ($-0.72$\,dex and $-2.36$\,dex for 47\,Tuc and M\,15, respectively, according to the catalogue of \citealt[][2010 edition]{harris1996}) and their small metallicity spread. On the other hand a large metallicity spread is already evident for the Phx dwarf galaxy.
The metallicity gradients and chemical enrichment of Phx is further discussed in Section~\ref{sec:metallicity}.

\section{Selection of Phoenix members} \label{sec:mem}

With Galactic coordinates $l\sim270^{\circ}, b\sim-79^{\circ}$, Phx is located in a region of the sky
that is not heavily contaminated by foreground stars. On the other hand, its systemic radial velocity
is not well separated from the line of sight velocity distribution of the MW disk.

The first pass of our Phx member selection was made on the basis of the stars' magnitudes and colours, retaining
targets whose location on the CMD
is consistent with RGB stars at the distance of the Phx dwarf galaxy (see Fig. \ref{fig:Phx-CMD}); this excluded 26 objects.

Since the radial velocities for the P71 FORS2 sample could not be reliably combined with the P83 sample,
the membership of these stars was further assessed solely on their
location in the FoV, [Fe/H] distribution, and the strength of the Mg I line at $8807$\,\AA, which is a proxy for surface gravity \citep{battaglia+starkenburg2012}.
Since this sample covers the innermost region of Phx, its metallicity distribution follows very well the metallicity distribution measured
from the P83 sample, and the EWs of the Mg I line are consistent with all stars being red giants, showing a smooth correlation with the EWs of the CaT lines with no obvious outliers, we consider that all these stars are Phx members.
In further support of this conclusion, stars of this sub-sample do not show a large velocity dispersion.

For the P83 sample, we excluded stars that have obviously outlying
radial velocities; then we applied further radial velocity cuts by inspecting the velocity gradient plot (see the bottom
left panel of Fig. \ref{fig:phx-rotation}) until only stars within three times the
observed (error inflated) velocity dispersion were left. The second step was
repeated iteratively several times until convergence.
The measurement of the EW of the Mg I line was possible for $135$ likely member stars. Its strength is consistent with all these stars being red giants within the measurement uncertainty according to the selection criteria proposed by \citet{battaglia+starkenburg2012}.

The sample of probable member stars so derived consists of 178 objects with both line-of-sight velocity and [Fe/H] measurements, and
further 18 objects with [Fe/H] measurements (for a total of 196 members).

To assess the remaining contamination of the sample,
we retrieved a Besan\c{c}on simulation of MW stars
\citep{robin+2003} in the observed field-of-view that have V magnitude
between 20.0 and 22.5 and colour 0.8 <V-I < 2.2. The Galactic model predicts $\sim$10 stars that would pass
all our selection criteria. This number can be regarded as an upper limit because the surface
number density of Phx is larger than the
surface density of MW contaminants at most radii, making
it more likely to pick a Phx member star when placing slits onto targets.

In Kacharov et al. (in prep.) we also determine membership probability on the
basis of the observed spatial distribution of Phoenix stars as well as the kinematic properties
predicted by axisymmetric Jeans mass modelling of the system, in presence of foreground contamination.
This approach predicts 191 stars with a probability of membership larger than $0.8$, in very good agreement with
the selection made by hard-cuts described above.

We have verified that the Phoenix properties presented in this article are unmodified by the choice of
one or the other approach for membership selection; the only change worth of notice is that the three most external stars
would not be classified as members in the Jeans modelling. In the following, we adopt the selection made on the hard-cuts
just for the sake of making the analysis independent on the details of the modelling. 

%\begin{figure}
%	\includegraphics[width=\columnwidth]{membership_probability}
%    \caption{Velocity - metallicity diagram for the full FORS2 MXU 2009 sample colour coded according to their membership probability in agreement with the best fit Jeans model.}
%    \label{fig:phx-rv-feh}
%\end{figure}

\section{Kinematic properties}
\label{sec:kinematics}

We used a maximum likelihood approach \citep{vandeven+2006,walker+2006} to compute the systemic heliocentric radial velocity of the Phx dwarf galaxy ($V_{helio} = -21.2\pm1.0$\kms) and its intrinsic velocity dispersion ($9.3\pm0.7$\kms, after correcting for the velocity gradient) taking into account the individual uncertainties of all the stars in the clean sample of probable member stars. The corresponding systemic velocity in the Galactic standard of rest is $V_{GSR} = -108.6\pm1.0$\kms\footnote{This is determined using the formula in \citet{binney+tremaine1987}, with a LSR velocity vLSR = 220 \kms at the distance
  of the Sun from the MW center 
  (R$_{\odot}$ = 8 kpc) and a solar motion of (U, V, W) = (10, 5.25, 7.17)\kms \citep{dehnen+binney1998}, where U is radially inward, V positive in the direction of Galactic rotation and W towards the North Galactic Pole. If we were to use  8.3 kpc from distance of the Sun from the Galactic centre 
  \citep{gwinn+1992}, a local rotation speed of $236$\kms \citep{bovy+2009} and a solar motion (U, V, W) = (11.1, 12.24, 7.25)
  \kms \citep{schoenrich+2010}, v$_{\rm GSR}$ would be $-114.2$\kms instead of $-108.6$\kms.}, i.e.\ Phx is approaching the MW.
At a distance of $\sim450$\,kpc, however, it has likely not yet encountered the hot Galactic halo extending to about $250 - 300$\,kpc and is moving through the Local Group medium, which ram pressure may be responsible for stripping of the remaining HI gas. Its velocity with respect to the Local Group centre of mass is $-100.8$\kms\footnote{The velocity with respect to the Local Group centre of mass is determined using the formula provided in \citet{tully+2008}.}.

The systemic velocity is consistent with the determination of $-13 \pm 9$\kms~ obtained by \citet{irwin+tolstoy2002} from a much smaller sample of 7 stars.
Worth noting is the excellent agreement of our measurements with the velocity and velocity dispersion of the HI cloud ($V_{\rm HI} = -23$\kms, $\sigma_{\rm HI} = 6$\kms) proposed by \citet{germain+1999} as physically associated to Phoenix.
The $-23$\kms cloud is more compact than other Galactic gas clouds along the line of sight \citep{germain+1999} and although not centred on Phoenix, its location and velocity structure are consistent with currently being stripped from the dwarf galaxy via ram pressure or SNe feedback from the last episode of star formation \citep{gallart+2001}.
However, we caution that at such low velocity the possibility of Milky Way gas contamination is still high.

It is well-known that dwarf galaxies inhabiting dense environments such as within
the virial radii of the MW or M31 are devoid of gas, except for relatively massive
systems such as e.g. the Magellanic Clouds or IC 10. The fact that such a small system
as Phoenix has been able to form stars until almost present day \citep{hidalgo+2009} 
and is likely to still contain gas, together with its large distance and approaching velocity, suggests that Phoenix is likely to
enter the MW virial radius for the first time and has evolved undisturbed from
interactions with large galaxies.
However, assuming a purely radial orbit and $10^{12}\,{\rm M_{\odot}}$ point mass for the MW, we estimate an orbital period in the order of $10$\,Gyr and an apocentric distance of $\sim1$\,Mpc.
Even on such an extended orbit, Phx has had time for at least one pericentric passage, hence velocity and distance alone cannot rule out the possibility of a close encounter with the MW in the past.

%We note that \citet{teyssier+2012} proposed Phoenix as one of the field Local Group galaxies likely to have at some point in time passed through the virial radius of the MW, based on the comparison between the
%GSR velocities and galactocentric distances of Local Group field galaxies with those of the populations of dark matter halos in the Via Lactea II (VLII) simulation with known orbital histories. This would appear in contradiction with our conclusion.
%However, their result relies on an adopted value of Phoenix systemic heliocentric velocity of 56\kms \citep{cote+1997}, which translates into v$_{\rm GSR} = -37$\kms.
%Our determinations remove this apparent contradiction and make Phoenix fully consistent with the properties of VLII sub-halos that have never entered the MW virial radius. 

In this section we make a detailed kinematic analysis of Phx using the same maximum likelihood technique to compute mean velocities and intrinsic dispersions for selected parts of the stellar sample.
Although we use the heliocentric radial velocities in the rest of the article, we confirmed that due to the small angular size of Phx, there is only a constant difference between the heliocentric and GSR velocities for all stars in the field with a standard deviation of just $0.15$\kms, which is too small to be accounted for in our sample.

\subsection{Rotation}

\begin{figure*}
	\includegraphics[width=16cm]{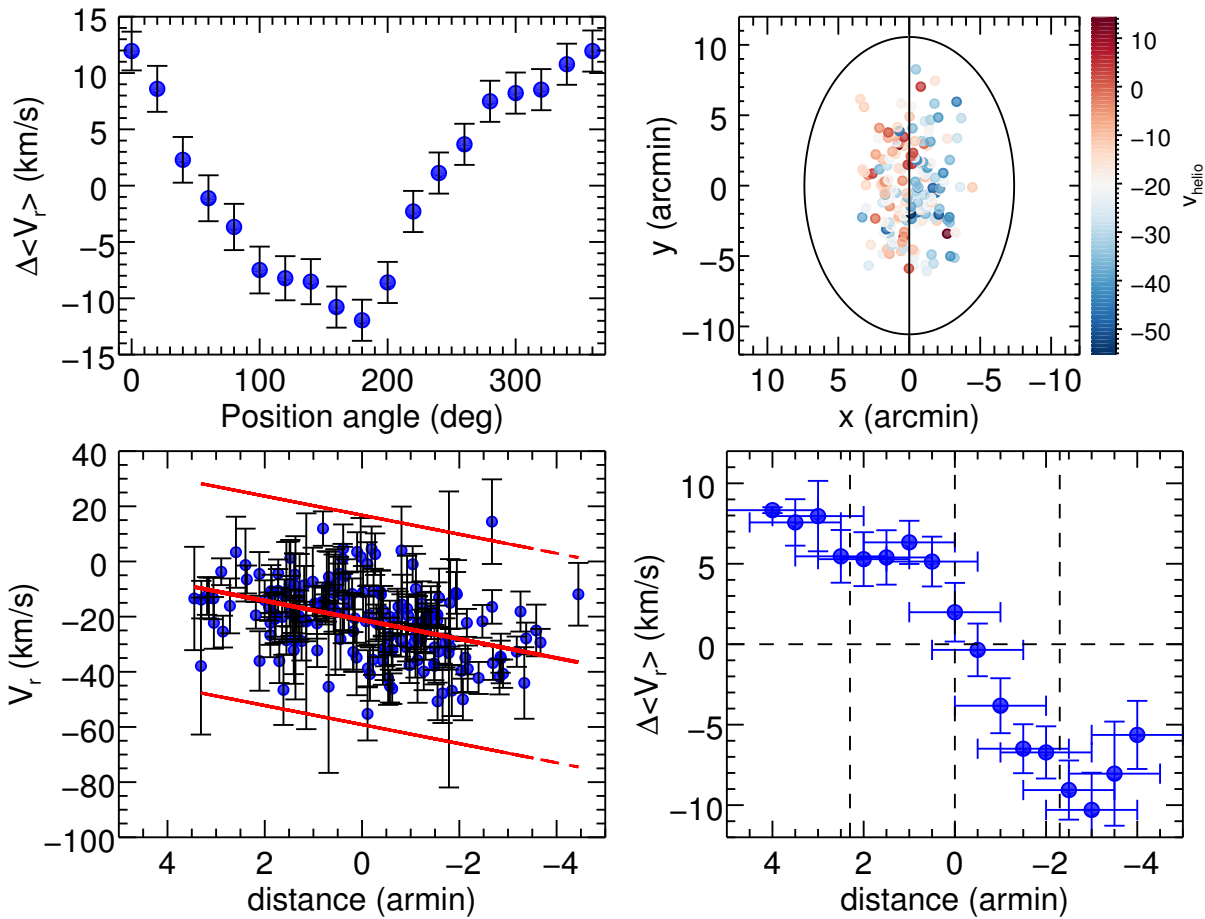}
    \caption{{\it Upper left panel:} Difference between the mean velocities on each side of the galaxy with respect to a line passing through its centre at a given position angle.
      {\it Upper right panel:} Spatial location of the observed Phx stars in a coordinate system where the axes are rotated to be aligned with the galaxy's major and minor axes.  The colour code denotes the line-of-sight velocity.
      The tidal ellipse \citep{battaglia+2012} and the adopted rotation axis are superimposed.
    {\it Bottom left panel:} Velocity gradient - the radial velocity of each probable member star with respect to a spatial axis perpendicular to the rotation axis of the galaxy. The red lines correspond to an error weighted linear fit to the data $\pm3$ times the intrinsic velocity dispersion.
    {\it Bottom right panel:} Rotation profile of Phx, where the abscissa shows the distance from the centre of the galaxy along the axis perpendicular to the rotation axis and the ordinate shows the mean offset from its systemic velocity in different overlapping bins. The horizontal error bars indicate the size of the selected bins, while the vertical error bars indicate the formal uncertainty of the mean velocity offset. The vertical dashed lines indicate the half-light radius of Phx.}
    \label{fig:phx-rotation}
\end{figure*}

We followed the procedure by \citet{walker+2006} to detect velocity gradients by measuring the difference between the mean velocities on either side of a line passing through the galaxy's centre and rotated at different position angles (Fig. \ref{fig:phx-rotation}; upper left panel).
Although the resulting curve is not symmetric there is a clear maximum at $0^{\circ}$ (minimum at $180^{\circ}$)
corresponding to a strong prolate rotation ($V_{rot}\sin i = 5.5\pm0.6$\kms, where $i$ is the unknown inclination angle). The velocity gradient can be clearly appreciated by eye as well e.g. on a plot showing the
radial velocities across the field-of-view of Phx (Fig. \ref{fig:phx-rotation}; upper right panel).

\begin{figure}
	\includegraphics[width=\columnwidth]{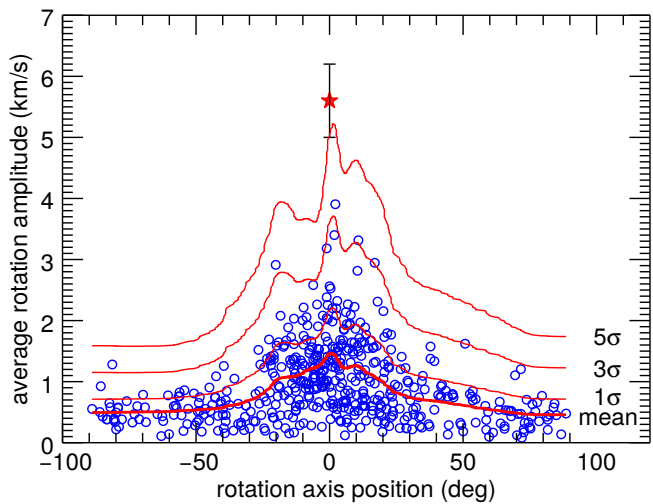}
    \caption{Monte Carlo realisation of false rotation detections. The plot shows the detected false rotation signal as a function of the rotation axis position ($0^{\circ}$ - prolate system, $\pm90^{\circ}$ - oblate system). The running mean is overplotted with a thick red line and the $1\sigma, 3\sigma, 5\sigma$ significance levels are overplotted with thin red lines. The detection in Phx is indicated by a red star.}
    \label{fig:rotation-significance}
\end{figure}

To further test the significance of the rotation detection, we performed a Monte Carlo test by simulating a mock galaxy that follows a 2D Sersic distribution with ellipticity of $0.3$ and contains $10^4$ stars.
This mock galaxy has a Gaussian velocity distribution with an intrinsic dispersion of $9$\kms and no intrinsic rotation.
We extracted the simulated stars that are closest to the observed $x,y$ positions in Phx and further reshuffled their mock velocities according to the velocity uncertainties measured in the observed sample at each given position.
Then we proceeded measuring the average rotation and rotation axis position angle of the resulting mock sample.
This was repeated $500$ times.
Fig. \ref{fig:rotation-significance} shows the estimated position angle vs.\ the average rotation amplitude of all Monte Carlo realisations.
Rotation axis of $0^{\circ}$ corresponds to a prolate rotator and $\pm90^{\circ}$ corresponds to an oblate rotator.
Indeed it is much more likely to detect a false prolate rotator than an oblate one.
Our detection of prolate rotation in Phx, however, is at $\ga 5\sigma$ significance level.

In the bottom left panel of Fig. \ref{fig:phx-rotation} we plot the radial velocities of the Phx probable member stars projected onto the axis of maximum velocity gradient (i.e.\ the axis perpendicular to the rotation axis). An error weighted linear fit is overplotted. That corresponds to a simple rigid body rotation model.
However, to better explore the rotation pattern of Phx, we construct a rotation profile in the bottom right panel of Fig. \ref{fig:phx-rotation} by plotting the difference between the systemic radial velocity of the galaxy and the mean velocity inferred from the stars in multiple overlapping bins along the axis of maximum velocity gradient.
We see that the mean velocity rises smoothly from the inner parts of this galaxy, starting to flatten out around $V_{max}\sin i\sim$8\kms
beyond the half-light radius \citep[$r_h = 2.3\arcmin$][]{battaglia+2012}.
Even though the exact shape cannot be pinned down with this data-set, it is evident that Phx displays a velocity gradient which is
aligned with its projected minor axis.
%We note that a variation in the P.A. of .... was also noticed in a proceeding
%by Zaggia et al. based on FLAMES/GIRAFFE data in Medusa mode. The detection of a similar behaviour in a different data-set and
%in a completely independent analysis lends further support to our findings. 

%Given that Phoenix is unlikely to have experienced a pericentric passage around the MW, we can most likely exclude possible tidal disturbance from our Galaxy as the cause of the observed velocity gradient.

Velocity gradients detected in objects with a large angular extent, such as
MW satellites, can arise due to the projection of the 3D systemic motion of the galaxy onto the line-of-sight towards the individual stars;
such gradients would mimic solid body rotation.
We explore what tangential motion could give rise to the velocity gradient we observe for Phoenix. 
Following the formalism by \citet{strigari+2010}, let us consider a Cartesian reference frame with the z-axis pointing towards the observer,
the x-axis positive in the direction of decreasing right ascension and the y-axis in the direction of increasing declination. In the small
angle approximation, the 3D systemic motion of the galaxy projects onto the line-of-sight of each star as
\begin{equation}
u = v_x \times (x/D) + v_y \times (y/D) - v_z \times (1 - (R^2/D^2)/2. )
\end{equation}
where $R = \sqrt{x^2 + y^2}$, $D$ is the distance to the galaxy, and $v_x$, $v_y$ and $v_z$ are the galaxy's systemic
velocity components. In the case of Phx, we trace the velocity gradient out to $R \sim 4$\arcmin\, approximately along the right ascension axis;
the term $R^2/D^2$ and $v_y$ can be therefore neglected and the above equation reduces to $u = v_x \times (x/D)$.
If we take the mean velocity $u$ at $x \sim$ 4\arcmin\, to deviate  
$\sim$5 to 8\kms\, from the systemic heliocentric velocity $v_z$, then this would correspond to a tangential motion $v_x$ between 4000 and 6600\kms\,!
Such large values are hardly reconcilable with the velocity dispersion of the Local Group, as well as with 
expectations for sub-halos found within $\sim$1\,Mpc from MW-sized halos in the
Aquarius simulations as well as Via Lactea-II and GHALO, be them
bound or unbound to the main halo \citep[e.g.\ see analysis of][]{Boylan-Kolchin+2013}. Hence the detected velocity gradient is not
to be attributed to perspective effects, which could account to about 1/10 of the detected amplitude.

%in order to calculate vgsr I used vsun=220kms, rsun=8kpc and the old sun motion; Teyssier et al.2012 used vsun=236kms, rsun=8.5kpc and
%The relative motion of the Sun is  = (11.1, 12.24, 7.25) \kms (Sch\onrich, Binney & Dehnen 2010).
%since vGSR= vh + $
%   vLSR * cos(gb_rad) * sin(gl_rad) + $
%   v_sm_mod * (cos(gb_rad) * cos_sm * cos((gl - l_sm) * degra) + sin(gb_rad) * sin_sm)
%for a system at a given l and b, vGSR and vh differ only by a constant.
% Teyssier et al. 2012 adopted  vh= 56kms and vgsr= -37kms. Hence  the term
%   vLSR * cos(gb_rad) * sin(gl_rad) + $
%   v_sm_mod * (cos(gb_rad) * cos_sm * cos((gl - l_sm) * degra) + sin(gb_rad) * sin_sm)
% for them is equal to  vGSR - vh = -37 - 56 = -93 kms.
% For me, Phx vgsr would then be  - 21.2 - 93 = -114.2  (only 6 kms different from my determination)

We conclude that the most likely possibility is that Phoenix displays prolate rotation, with a rotation-to-dispersion ratio
approaching 1 in the outer parts. The clear signal of prolate rotation in Phx is puzzling from a dynamical point of view. Prolate rotation has been occasionally observed in giant elliptical galaxies, where the morphology is determined by the shape of a triaxial dark matter halo \citep[e.g.][]{schechter+gunn1978,emsellem+2004,vandenbosch+2008}, while most dSphs have been found to be oblate.
An interesting case is the Andromeda II dSph, which to our knowledge
is the only other example of a prolate rotating dwarf galaxy \citep{ho+2012}. \citet{amorisco+2014}
detected a cold stellar stream in the field of this galaxy and suggested that a merger event has significantly altered its kinematics.
\citet{fouquet+2016} provide further evidence from N-body and hydrodynamical simulations that a major merger was responsible for the prolate rotation in And II. It is possible that Phoenix evolution mostly unaffected by interactions with the MW may have allowed it to experience a major merger with another dwarf - something extremely unlikely in the vicinity of a giant galaxy \citep{tremaine1981,derijcke+2004}.
%In that scenario dwarf-dwarf mergers might be one of the missing evolutionary links that connect 
%dIrr and dSph galaxies \citep[see also][]{benitez-llambay+2016}.

It is intriguing to note that the spatial distribution of young stars in Phoenix is clearly tilted by
$\sim90^{\circ}$ with respect to Phx older stellar component \citep{MD+1999, hidalgo+2009, battaglia+2012},
i.e. it is aligned with the projected minor axis of the main body
and the direction of the maximum velocity gradient. It is possible that the events that caused the prolate rotation are also
responsible for the peculiar distribution of the recently formed stars. 

An alternative possibility for the appearance of this odd velocity gradient may be sought if Phx did experience one close encounter with the MW and displays clues of tidal deformation.

In the upcoming paper (Kacharov et al., in prep.) we will examine both the rotation signal and the velocity dispersion profile
via orbit modelling and Jeans modelling of Phoenix as a prolate, axisymmetric body. We defer Phx mass estimates, as well as the
presentation of the velocity dispersion profile to that paper. 

%{\bf examine 2D distribution of rotation field subtracted l.o.s. velocities. Do we see anything? A stream?

%  re-examine displaced clump of stars that were seen in battaglia2012; connected to this possible accretion event?
%}

\section{Metallicity properties}
\label{sec:metallicity}

\begin{figure*}
	\includegraphics[width=16cm]{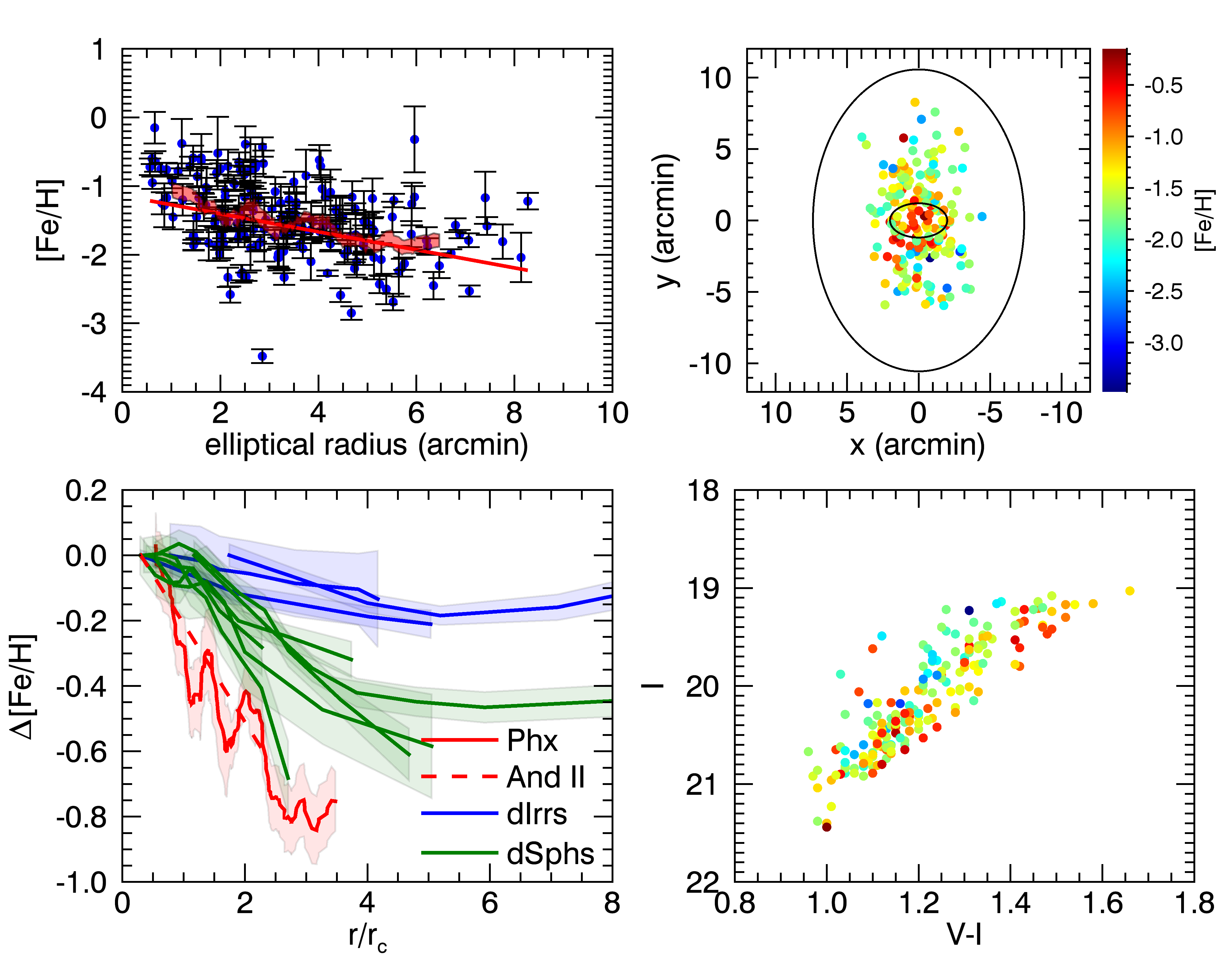}
        \caption{{\it Upper left panel:} Metallicity gradient in the Phx galaxy - the metallicity of each star as a function of its elliptical radius
          is shown by the circles with error-bars. An error weighted linear fit to the data is indicated with a red line and an error weighted, running mean computed at the position of each star based on the 20 stars with closest elliptical radial coordinates is indicated with a transparent, red curve. The thickness of the curve corresponds to the uncertainty of the mean.
          {\it Upper right panel:} Spatial extent of the observed Phx probable member stars. The colour code denotes the metallicity. The outermost and innermost ellipse indicate
          the nominal King tidal radius and the approximate location of the young stars, respectively \citep{battaglia+2012}.
          {\it Bottom left panel:} The metallicity gradient of Phx (red) is compared to those of MW dSphs (green) and
          gas-rich objects (WLM, SMC, LMC, in blue) as presented in \citep{leaman+2013}. The metallicity gradient of And II \citep[red dashed line][]{ho+2015} is also presented assuming a core radius of $5.2$\,arcmin \citep{mcconnachie+irwin2006}.
          {\it Bottom right panel:} Location of Phx member stars on the colour-magnitude diagram: a clear colour-metallicity correlation is visible.
    }
    \label{fig:phx-metallicity}
\end{figure*}

We find that Phx is a relatively metal poor system with a large intrinsic metallicity spread: i.e.\ by approximating the metallicity distribution function (MDF) with a Gaussian, the maximum likelihood value for the mean metallicity of the observed sample is $\rm{[Fe/H]}=-1.49\pm0.04$\,dex with an intrinsic dispersion $\sigma_{\rm{[Fe/H]}} = 0.51\pm0.04$\,dex.

\subsection{Spatial variations} \label{sec:metgra}

In the upper left panel of Fig.~\ref{fig:phx-metallicity} we show the metallicity of each probable member star as a function of elliptical radial distance from the galaxy's centre.
There is a clear metallicity gradient with more metal rich stars located in the central regions and more metal poor stars
spread out at all radii. An error weighted linear fit suggests a central metallicity of $-1.14\pm0.02$\,dex, which gradually decreases by $0.13\pm0.01$\,dex\,arcmin$^{-1}$ ($0.23$\,dex per $r_c$); the trend is confirmed by a running average.
The upper right panel of Fig.~\ref{fig:phx-metallicity} shows the 2D metallicity distribution of our sample of Phx member stars.
It seems that the most metal rich giant stars (representative of the intermediate age population, see Sect.~\ref{sec:agemet}) are aligned with the central bar-like feature of young main sequence stars \citep[see][]{MD+1999, hidalgo+2009, battaglia+2012} and indicated by the position of the small, inner ellipse in the figure.
This radial metallicity gradient is fully in agreement with the radial star formation history (SFH) gradient discussed by \citet{hidalgo+2009}, according to which young an intermediate age stars, which are the most metal rich ones, are concentrated in the central part, gradually disappearing at large galactocentric radii. Thus, the origin of the metallicity gradient is actually to be found in the age gradient present in the galaxy, and in the existence of an age-metallicity relation. The gradient would be much flatter if only older stars were considered.

Negative metallicity gradients of varying steepness have been observed
in most MW and in some M\,31 dwarf spheroidal galaxies of comparable luminosity to Phx.
It is remarkable that the only other Local Group dSph with prolate rotation discovered so far, And~II, has a metallicity gradient of similar steepness as Phx, and both are the systems in which the steepest gradient has been measured.
On the contrary, there are indications that shallower gradients are present in more massive systems \citep[Fig.~\ref{fig:phx-metallicity}, bottom left, see also Fig.~10 in][]{ho+2015}.
The fact that such steep metallicity gradients are present in low-mass dwarf galaxies inhabiting a dense environment (i.e. satellites of the MW or M\,31) and in a system found beyond the Galactic virial radius and possibly at its first approach towards the Galaxy might point to metallicity gradients being an intrinsic property of these small galaxies, possibly linked to the age gradients generally found in them (and very pronounced, particularly, in Phx and And II).
The result appears slightly more difficult to reconcile with the hypothesis that a higher rotational-vs-pressure support produces shallower metallicity gradients \citep{leaman+2013}, given the larger average rotation of Phx (and And II) with respect to classical MW dSphs. If however, the prolate rotation in these systems was caused by an accretion or merger event, it is not excluded that Phx or And II rotation might have been lower in the past, bringing them in better agreement with this hypothesis. 

%It is interesting to see that Phx displays the steepest metallicity gradient among the dwarfs analyzed so far.
The steep metallicity gradient of Phx is unlikely to be an artifact, as metallicities were determined using the same spectral lines and calibration as
for most of the objects analyzed in \citet{leaman+2013}.
Since Phx was able to form stars until very recently, while star formation in the majority of MW dSphs stopped long ago ($\gtrsim7-8$\,Gyr ago),
it is reasonable to speculate that the HI gas had time to concentrate further in the center of Phx
\citep[provided angular momentum was not acting as a ``centrifugal barrier'', see][]{schroyen+2011}.
From the age-metallicity relation derived in Sect.~\ref{sec:agemet}, it is evident that the more metal-rich stars, which are found within
the inner $3$\,arcmin, are the youngest among the targeted
RGB stars. If star formation in Phx would have stopped a few Gyr ago, it is very likely that we would have observed a shallower gradient. 

In the bottom right panel of Fig.~\ref{fig:phx-metallicity} we show the location of Phx members on the CMD. A colour-metallicity dependence is clearly detectable, although slightly smeared due to the age-metallicity degeneracy.
The metal poor stars are found predominantly at bluer colours, while the more metal rich stars are found at redder colours.
%{\bf GB: this last panel is nice, but I am not sure if it is needed. Nikolay, can you add a few words in the text as to why we are showing it?}

\subsection{Age-metallicity relation}
\label{sec:agemet}

\begin{figure}
	\includegraphics[width=\columnwidth]{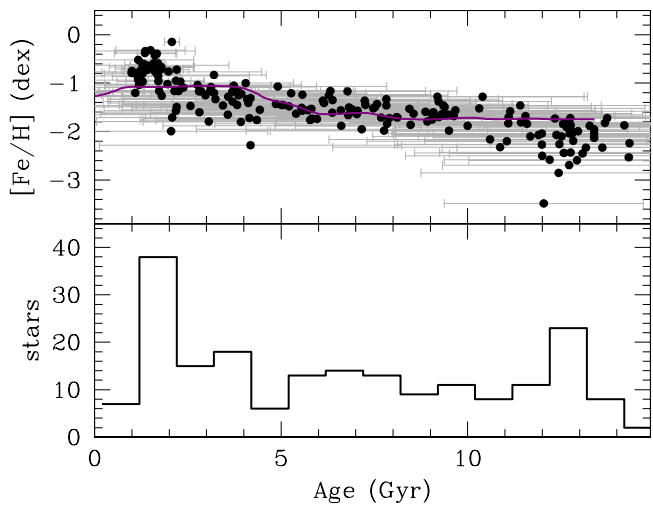}
        \caption{Age-metallicity relation for Phx stars in our VLT/FORS2 MXU sample, derived using
          Eq.~\ref{eq_age} (top). The bottom panel shows the age distribution.
          The purple solid line indicates the age-metallicity relation derived by \citep{hidalgo+2009}
          from HST CMD diagrams.
    }
    \label{fig:amr}
\end{figure}

The spatially resolved star formation and chemical evolution history of Phoenix were obtained by \citet{hidalgo+2009} from
deep HST/WFPC2 colour-magnitude diagrams reaching out to $\sim$5\,arcmin from the centre.
The age-metallicity relation shows
the metallicity to be approximately constant around [M/H]$=-1.7$ in the first 7 Gyr of evolution, 
to then increase more quickly to [M/H]$\sim -1.1$ and
finally flatten around that value or even decrease (Fig.~\ref{fig:amr}).

Deriving stellar ages for RGB stars is hampered by the well known age-metallicity degeneracy and carries large uncertainties.
However, we can attempt to estimate relative stellar ages of our sample of RGB stars since we have obtained
spectroscopic metallicities from the strength of the CaT lines, hence 
independently from the CMD information.

Here we use the relation by \citet{carrera+2008a} to infer ages for the stars in our sample. 
We refer the reader to the original work for details; in summary,
from a synthetic CMD created with a SFH and chemical enrichment law sampling a broad age and metallicity
range (13 $\le$ age [Gyr] $<$0 and $-2.5 \le$ [Fe/H] $\le +0.5$), the authors obtained a general relation between age, dereddened 
$(V-I)_0$ colour, V-band absolute
magnitude $M_V$, and metallicity for RGB and AGB stars fainter than the RGB tip\footnote{The synthetic CMD was created with IAC-star \citep{aparicio+gallart2004} and the BaSTi stellar evolution library \citep{pietrinferni+2004}.}:  
\begin{equation}\label{eq_age}
\begin{split}
log_{10}(\mathrm{age})=2.57+9.72\times(V-I)_0 +0.70\times M_V -1.51\times \mathrm{[Fe/H]} + \\
-3.86\times(V-I)_0^2-0.19\times \mathrm{[Fe/H]}^2+0.49\times(V-I)_0^3
\end{split}
\end{equation}
The distance modulus and reddening we used to obtain absolute magnitudes and correct \vi~colours are listed in 
Table~\ref{tab:parameters}.

%The first one consists on deriving the age of each star by comparing its location on the CMD with the set of isochrones from the
%Yonsei-Yale library \citep{yi+2001, kim+2002} of the appropriate metallicity.
For each star, we first computed $10^4$ random realisations of
its $(V-I)$ colour, $M_V$, and [Fe/H], extracting them from 
a Gaussian probability distribution centered
on the observed values and with $\sigma$ corresponding to the measurement error for each parameter. 
We then obtain a distribution of ages, by applying Eq.~\ref{eq_age} to these $10^4$ random realisations, and 
adopt as age and 
its uncertainty for each star the mean and the standard deviation of the distribution. 
To avoid the regime of saturation of Eq.~\ref{eq_age}, which occurs at $\gtrsim$12.5 Gyr, as well as 
unrealistic ages such as younger than that of the youngest stars in the upper part of the
RGB, $\sim$ 1 Gyr, we assign a random value within the uncertainty to those objects with ages 
falling beyond these extremes. 

The metallicity as a function of age is shown in the top panel of Fig.~\ref{fig:amr}. 
%The errorbars denote the
% uncertainties in the age determination in each case. The main difference are observed in the edges because of the assumptions used for the youngest and oldest ages obtained. This is clearly obs
%erved in the bottom panel of Fig.~\ref{fig:amr}. Whereas in the case of isochrones there is a buildup of stars in the edges, in the case of Equation~\ref{eq_age} the stars are more spread. This 
%is particularly clear in the case of the oldest ages. Anyway, in both cases it is observed a similar trend. 
The metallicity is found to be [Fe/H]$\sim-2.0$ dex at the oldest ages, $\gtrsim$12 Gyr ago, and, after a slight increase, it 
remains more or less constant at [Fe/H]$\sim -1.6$ until about 4 Gyr ago. Finally, a new increase of metallicity 
is observed in the last 2 Gyr, where [Fe/H] reaches up to $\sim-0.5$\,dex.  

Our findings are in fairly good agreement with the age-metallicity relation derived from the Phx SFH by 
\citet{hidalgo+2009}, of which we reproduce the main traits in Fig. \ref{fig:amr}. In the latter, 
the increase of metallicity at the beginning of the galaxy life is not observed,  
possibly because of the lack of age resolution in the SFH determination at these old ages.
The colour of RGB stars scales linearly with $Z$ and the isochrones become degenerate in metallicity at low [Fe/H].
Between $\sim 9$ and $5$\,Gyr ago, the agreement of the age-metallicity relation derived from the CMD with the spectroscopic measurements is excellent.
%Other noticeable differences between the photometric and spectroscopic age-metallicity relations are observed for ages younger than $4$\,Gyr, since in the photometric case the metallicity remains constant around [Fe/H]$\sim-1$, while in the spectroscopic case one can notice a clear increase.
The differences observed for ages younger than $4$\,Gyr follow from the fact that an upper limit for the metallicity of [Fe/H]$\sim-1$ was assumed in the SFH derivation by \citet{hidalgo+2009}, and thus the flat profile at young ages in the photometric case, while in the spectroscopic determinations one can notice a clear increase.

It should be noted that Eq.~\ref{eq_age} was derived assuming a ``MW-like'' trend for [$\alpha$/Fe] versus [Fe/H]:
\begin{equation}
[\alpha/Fe]=\left\lbrace\begin{array}{cl}
0.4 & \mbox{if [Fe/H]$\leq$-1.0}\\
-0.4\times[Fe/H] & \mbox{if [Fe/H]$>$-1.0}
\end{array} \right.
\end{equation}
which needs not be appropriate for a dwarf galaxy system in the mass range of
Phx, where the ``knee'' of [$\alpha$/Fe] is likely to occur at much lower metallicities than for the MW
\citep[see][for examples of the different trends exhibited by dwarf galaxies]{tolstoy+2009}.
On the other hand, the photometric age-metallicity relation was obtained 
by using solar-scaled models, while a plateau of $\alpha$-enhancement at low metallicities
might have been expected. While the actual chemical properties of Phx are likely to lie somewhere in between these
assumptions, it is reasonable to think that the observed differences might partially be due
to the adopted [$\alpha$/Fe] trends. Hence, overall we deem the comparison satisfactory.

The lower panel of Fig. \ref{fig:amr} shows the age distribution inferred for the stars in our spectroscopic sample. Note that this age distribution cannot be directly compared with the SFH of the galaxy, as derived for example by \citet{hidalgo+2009}. The SFH, by definition, provides the information on the amount of mass transformed into stars at a given time, while the fraction of this mass present in alive stars of given age decreases strongly toward older ages.
In a random sample of RGB stars near the RGB tip, the number of old stars is underestimated relative to the number of young stars by a few tens of percent due to the varying lifetime of RGB stars as a function of mass \citep{cole+2009}.
Thus, the true age distribution is likely to be suppressed at the young end and enhanced at the old end compared to the shown histogram.

%In order to test the age determination method here used and
%quantify the effect of different [$\alpha$/Fe] trends, 
%we also attempt age determination via a simple isochrone fitting
%(middle panel Fig.~\ref{fig:amr}); to this aim, we use the Yonsei-Yale library \citep{yi2001, kim2002} and test both
%the [$\alpha$/Fe] trend described above and
%runs of constant [$\alpha$/Fe](=0.0, +0.3).

\subsection{Discrete analytic chemical enrichment models}

Modelling the MDF of dwarf galaxies has significantly increased our understanding of
their chemical enrichment: e.g.\ the shape of the MDF of MW dSph satellites strongly points to
a loss of 96\% to $>$99\% of the metals they have produced \citep[e.g.][]{kirby+2011b}.
% {\bf should quote matteucci's works?}.
%The analysis of the linear metallicity distribution of Local Group dwarf galaxies and stellar clusters have been found to be binomial in form, which can be explained with a simple inhomogeneous chemical evolution model in terms of the number of enrichment events, covering fraction, and intrinsic size of the enriched regions \citep{leaman2012}. 

In order to interpret the metallicity properties of a galaxy in terms of its chemical evolution, we need to retrieve the shape of its {\it global}
MDF. In systems which display metallicity gradients, the observed MDF may suffer from biases, depending on the details of the observing strategy.
In the specific case of our slit mask observations of
Phoenix, the spatial distribution of the spectroscopically observed member stars is relatively uniform with radius,
unlike the underlying stellar population that is best described by a
Sersic profile \citep{battaglia+2012}. Coupled to the observed 
strong metallicity gradient, this may cause our sample to miss many of the centrally concentrated metal rich stars, while over representing the more metal poor stars,
which have a more extended spatial distribution. We note that the opposite might happen in systems with strong metallicity gradients where only the
central regions have been surveyed. 

In this work we adopt the \citet{kirby+2011} analytic models to investigate
how inflow and outflow of gas have affected the star formation in Phx, but we introduce few important modifications
to account for the spatial sampling of the spectroscopic data set with respect to the underlying  stellar density.

First we consider the pristine ({\it leaky box}) model, which has no gas inflow and where
the SNe yields and gas outflow are controlled by a single parameter - the effective yield ($p$).

We then include the dependence on the elliptical radial distance from the centre of the galaxy, $r$,
by considering the two-dimensional probability distribution function: 
\begin{equation}\label{eq:2D_MDF}
 P(\rm{[Fe/H]},r) \propto 10^{\rm{[Fe/H]}} \exp\bigg(-\frac{10^{\rm{[Fe/H]}}}{p(r)}\bigg).
\end{equation}
The effective yield parameter in units of $\rm{Z/Z_{\odot}}$ is radially dependent, assuming a linear dependence in [Fe/H] space (exponential in $\rm{Z}$ space):
\begin{equation}\label{eq:grad}
 p(r) = p_0 10^{r\Delta\rm{[Fe/H]}},
\end{equation}
where $p_0$ is the central effective yield and $\Delta$[Fe/H] is the metallicity gradient.
The radial dependence of the effective yield parameter could be interpreted as the SNe yields being better retained in the central regions of the galaxy and more efficiently expelled from the outer parts, where the gravitational potential is shallower.
\citet{schroyen+2013} showed that metallicity gradients are robust over time in dwarf galaxies, where star formation proceeds in more and more central regions as time passes, and are not destroyed via radial migration of stars, as happens in large spiral galaxies for instance.

The model with gas inflow was first introduced by \citet{lynden-bell1975} and also used by \citet{kirby+2011} to show that it describes well the central MDF of some dSphs (e.g. Fornax, Leo I, Draco).
In this framework the galaxy continuously accretes decreasing amount of metal free gas that is available for star formation.
Additionally to the effective yield, this family of models is also characterised by a second parameter ($M > 1$), which describes the accreted gas mass.
If $g$ is the mass of the available gas in the galaxy as a fraction of the initial gas mass and $s$ is the stellar mass in the same units then:
\begin{equation}
 g(s) = \bigg(1-\frac{s}{M}\bigg) \bigg(1+s - \frac{s}{M}\bigg)
\end{equation}
In the ideal case, where all the gas is converted into stars, $M$ is equal to the final stellar mass as a fraction of the initial gas mass.
In the case when $M=1$, there is no extra gas and the model reduces to the pristine case.
The only modification to the probability distribution function presented in \citet{kirby+2011} is that we again assumed radially dependent effective yield in the same form as in Eq. \ref{eq:grad}.
The extra gas model in our case has 3 free parameters: $p_0$, $\Delta$[Fe/H], and $M$.
The MDF is given by the following two equations:
\begin{equation}\label{eq:extra-gas1}
\begin{split}
  {\rm [Fe/H]}(s) = & \log\Bigg(p(r)\bigg(\frac{M}{1+s-\frac{s}{M}}\bigg)^2 \\
  & \times \, \bigg(\ln\frac{1}{1-\frac{s}{M}} - \frac{s}{M}\Big(1-\frac{1}{M}\Big)\bigg)\Bigg)
\end{split}
\end{equation}
\begin{equation}\label{eq:extra-gas2}
 P({\rm [Fe/H]},r) \propto \frac{10^{\rm [Fe/H]}}{p(r)} \frac{1+s(1-\frac{1}{M})}{(1-\frac{s}{M})^{-1} - 2(1-\frac{1}{M})\times10^{\rm [Fe/H]}/p(r)},
\end{equation}
where we solve Eq. \ref{eq:extra-gas1} numerically for $s$ and use the solution in Eq. \ref{eq:extra-gas2}.

Our likelihood function takes the form:
\begin{equation}
\begin{split}
 \mathscr{L}(p_0, \Delta{\rm[Fe/H]}, (M) \, | \, {\rm[Fe/H]}_i, \delta{\rm[Fe/H]}_i, r_i) = \\
 \prod_i \int_{-\infty}^{+\infty} P({\rm[Fe/H]}, r_i) \, \times \, S(r_i) \, \times \, r_i \\
 \times \frac{1}{\sqrt{2\pi}\,\delta{\rm[Fe/H]}_i} \, \exp \Bigg(-\frac{({\rm[Fe/H]}-{\rm[Fe/H]}_i)^2}{2(\delta{\rm[Fe/H]}_i)^2}\Bigg) \, d{\rm[Fe/H]},
\end{split}
\end{equation}
where $S(r_i)$ is the Sersic probability at the position of each star and $\rm{[Fe/H]_i}$ and $\delta\rm{[Fe/H]_i}$ denote the metallicity and the metallicity uncertainty of each star, respectively.
The parameters of the Sersic profile that best describes the galaxy are fixed from the photometric stellar density profile (see Table \ref{tab:parameters}).
%( ($\alpha_0$, $\delta_0$)=  27.77642deg, -44.44469deg, Sersic profile parameters  1.67 ;arcmin (sersic radius)
%  m = 0.88 ; (sersic power)
%  eps = 0.3 ; (ellipticity)
%  theta = 5. ;degrees (ellipse position angle)

To find the best set of parameters we maximize the log-likelihood function ($\hat{\mathscr{L}} = \log \mathscr{L}$) using the Metropolis-Hastings Markov Chain Monte Carlo (MCMC) algorithm \citep{hastings70}.

\begin{figure*}
	\includegraphics[width=14cm]{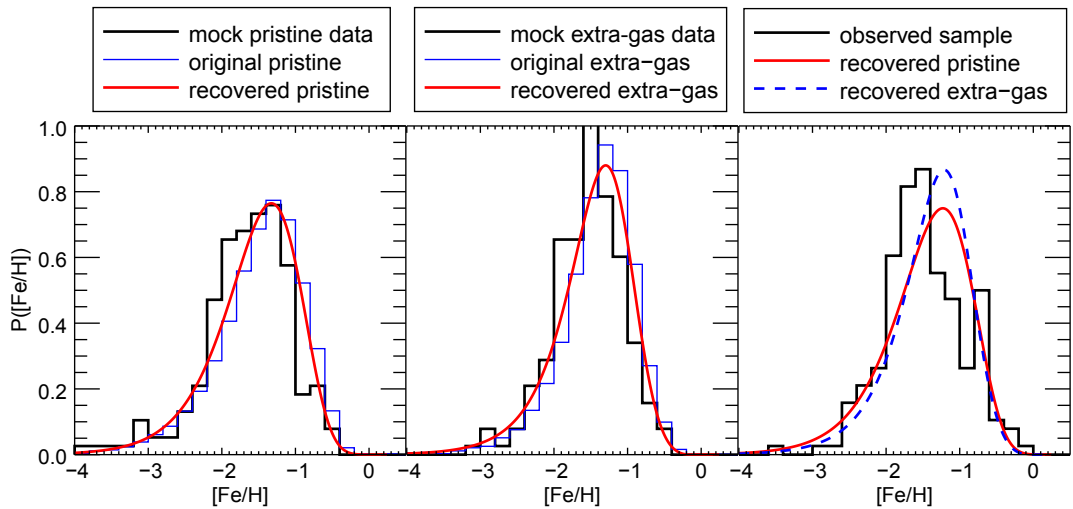}
    \caption{Example mock data and recovered MDF for the pristine ({\it left panel}) and the extra gas ({\it middle panel}) models. The blue histogram represents the complete mock-galaxy sample, the black histogram represents the mock observations sample, which was used to recover the original MDF - red line; {\it right panel:} the observed MDF of Phx - black histogram. The recovered full MDFs according to the pristine and extra gas model are overplotted with red and blue-dashed lines, respectively.
    }
    \label{fig:mdf}
\end{figure*}

\begin{figure}
	\includegraphics[width=\columnwidth]{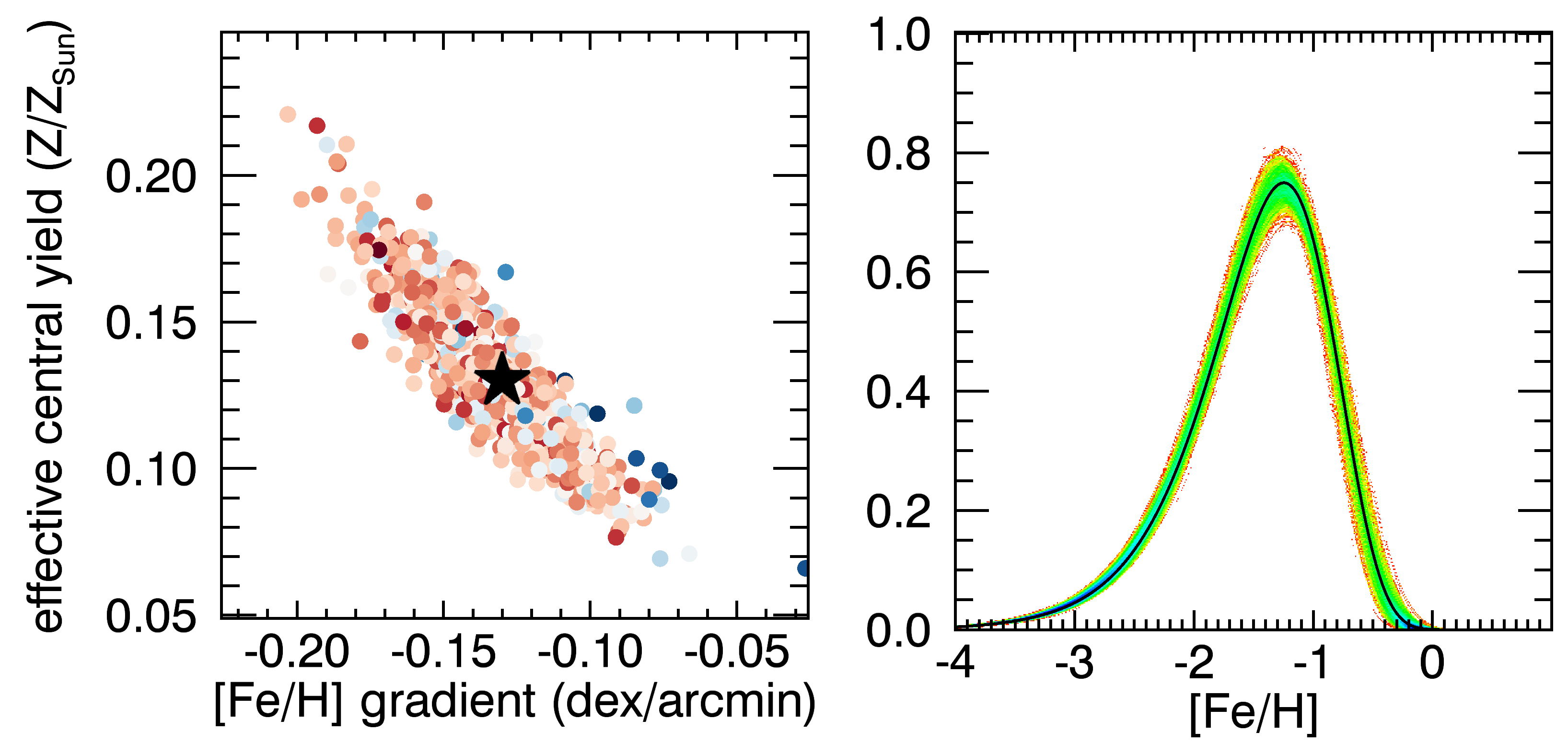}
    \caption{The recovery of the original MDF of 1000 mock data samples taken from the pristine model. The recovered parameters (metallicity gradient and central effective yield) are plotted in the left panel colour coded by maximum likelihood found during the fit (red - higher likelihood, blue - lower likelihood). The black star indicates the original parameters used to generate the mock data. The resulting MDFs are presented in the right panel colour coded by density. The original MDF is presented with a thick black line.
    }
    \label{fig:pristine}
\end{figure}

\begin{figure*}
	\includegraphics[width=14cm]{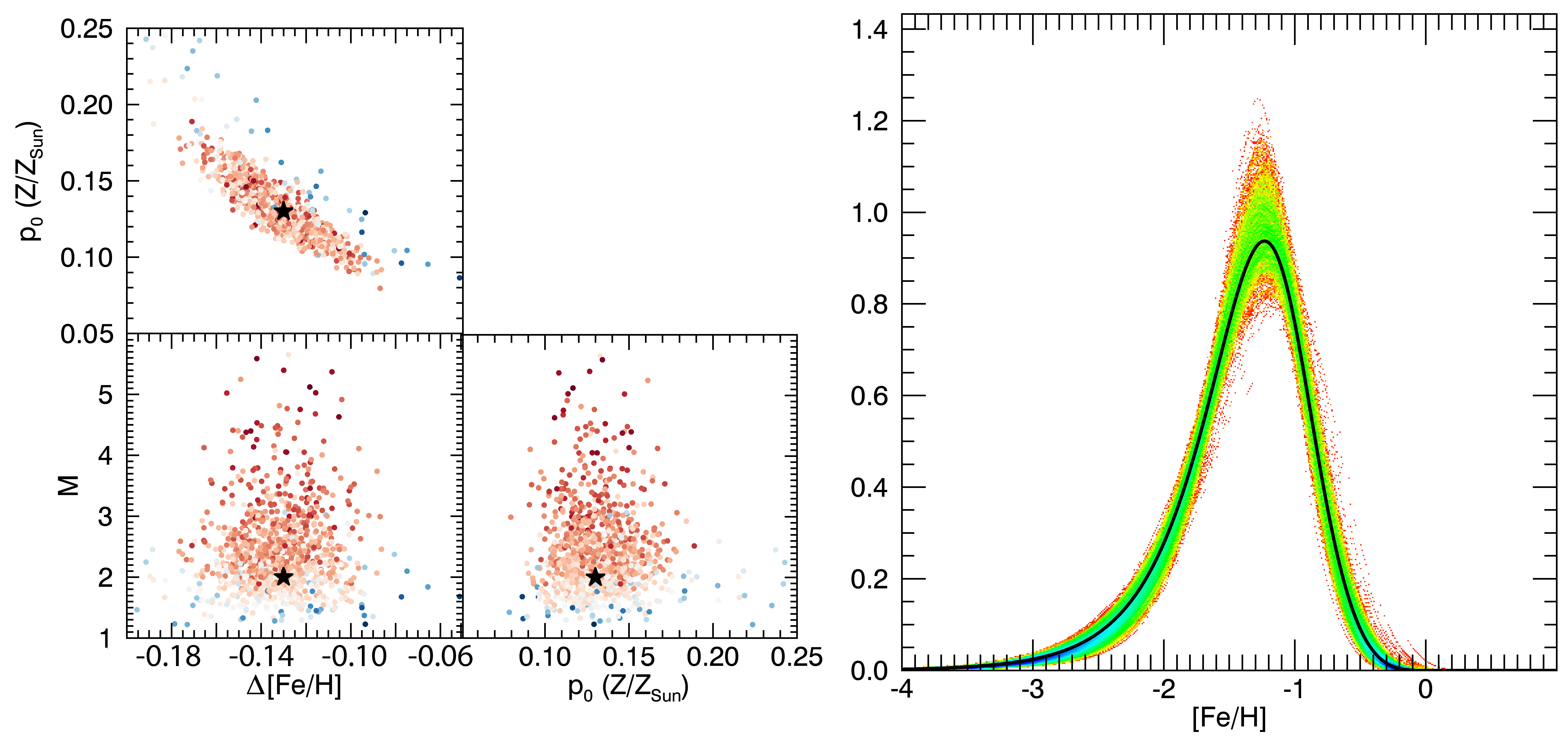}
    \caption{The recovery of the original MDF of 1000 mock data samples taken from the extra gas model. The recovered parameters ($\Delta$[Fe/H], $p_0$, $M$) are plotted in the left panel colour coded by maximum likelihood found during the fit (red - higher likelihood, blue - lower likelihood). The black stars indicate the original parameters used to generate the mock data. The resulting MDFs are presented in the right panel colour coded by density. The original MDF is presented with a thick black line.
    }
    \label{fig:extra-gas}
\end{figure*}

Before turning to the observations, we tested the ability of our method to recover the global MDF
by simulating mock data of similar size and spatial sampling as for our Phoenix
spectroscopic sample. This is illustrated in Fig. \ref{fig:mdf}.
The creation of the mock data samples is as follows: first we create a mock galaxy with $10^4$ stars that follow a Sersic distribution; then we randomly assign a metallicity to each star drawn from our 2-dimensional probability distribution function (Eq. \ref{eq:2D_MDF} in the case of the pristine model).
The metallicity gradient and effective yield parameters were chosen to be similar to what we find in Phx
(see below), but in the extra gas case, we decided to use $M = 2$ to make it distinct from the pristine model. 
The global MDF of the ``mock galaxy'' is presented with the blue histogram in Fig. \ref{fig:mdf}.

Then we select the stars with positions closest to the positions of the observed Phx stars and assign to them a metallicity uncertainty in agreement with the observed [Fe/H] error distribution (Fig. \ref{fig:snr-v-feh}), mimicking our target selection in spectroscopic masks.
The resulting MDFs (black histograms in the left and middle panels of Fig. \ref{fig:mdf}) are qualitatively similar to the observed MDF of Phx
(see the right panel of Fig. \ref{fig:mdf}).
The recovered MDF (red lines) are in excellent agreement with the original distributions for both the pristine and extra gas models,
notwithstanding the spatially biased and size limited mock data samples. 
We made $1000$ random realisations of the mock samples as drawn from the pristine and the extra gas models and tried to reconstruct the original distributions.
The recovered parameters and resulting MDFs are presented in Fig. \ref{fig:pristine} for the pristine model and
Fig. \ref{fig:extra-gas} for the extra gas models, respectively.
The central effective yield and metallicity gradient parameters appear to be correlated in both models but well centred around the true values.
Nonetheless, the trend of metallicity vs.\ radius (Fig. \ref{fig:phx-metallicity}) could be used for a direct determination of the
metallicity gradient, breaking the degeneracy among the model parameters. In the case of the extra gas model, however, a limited sample of $\sim190$ stars 
with fairly large [Fe/H] uncertainties does not constrain well enough the $M$ parameter, in general leading to an overestimate.
We confirmed that if larger mock samples ($\sim1000$ stars) with reduced [Fe/H] errors are used, then the estimate of the $M$ parameter agrees better with the
input value, although is still slightly biased towards higher values. 

%{\bf I would really try to use the direct information on the metallicity gradient to break the degeneracy between gradient and effective yield, and possibly improve the determination of M. I am not too optimistic about M because it seems uncorrelated with the other parameters.}

Fitting the pristine model to the Phx observations we find $p_0=0.13\pm0.01$ and $\Delta{\rm [Fe/H]} = -0.13\pm0.01$\,dex per arcmin,
where the quoted uncertainties are the $1\,\sigma$ spreads of the parameters in the Markov chain after the burn-in phase.
The model estimate of the gradient slope is the same as inferred from the simple linear fit to the run of [Fe/H] as a function of radius
described in Sect.~\ref{sec:metgra}. In agreement with the pristine model, also with the extra-gas model we obtain
$p_0 = 0.14\pm0.01,\, \Delta{\rm [Fe/H]} = -0.13\pm0.01$; the parameter quantifying the fraction of accreted gas is $M = 1.6\pm0.2$;
considering that the extra gas parameter tends to be overestimated in our mock samples, we conclude that it is unlikely that
Phx accreted large amounts of gas throughout its evolution.
When seen in the context of our hypothesis of Phx prolate rotation as possibly
caused by a past accretion/merger event, this would imply that the accreted system should not have carried with it large amounts of gas.
However, our analysis does not include the youngest stellar population of Phx, which may have been born from a recently captured gas-rich system that is responsible for both the observed prolate rotation and the peculiar distribution of the young stars in a disk/bar like structure perpendicular to the morphology of the old population.

The resulting total MDFs from the two models are presented in the right panel of Fig. \ref{fig:mdf}.
In both cases, the recovered MDF peaks at slightly higher [Fe/H] than the observed one.
One should keep in mind, of course, that the reconstructed MDFs are only realistic if the galaxy does indeed follow one of those empirical enrichment models.
While \citet{kirby+2011} did show that some dSphs' MDFs can be well approximated with these models, there are also notable exceptions in their sample
like Sculptor and Leo II. We also attempted to correct the observed MDF for spatial sampling without having to assume a model for the MDF, simply
by assigning a weight to each spectroscopic member as in \citet{larsen+2012}, i.e. calculating the ratio between the number of stars predicted by the Sersic model and spectroscopically observed per bins of elliptical radius.
Reassuringly, the resulting MDF shares the same features as those reconstructed from the analytical models, in what the contribution of the metal-poor tail decreases and
the [Fe/H] value of the peak increases with respect to the observed MDF, although it is considerably noisier because of the relatively low number statistics.  

\citet{kirby+2013} used the weighted mean [Fe/H] to place galaxies on the luminosity-metallicity relation.
Assuming a stellar mass $M_* = 0.77 \times 10^6$ \sm \citep[][and references therein]{mcconnachie2012}, Phoenix fits
well in the \citet{kirby+2013} luminosity-metallicity relation, both 
when we use a mean metallicity $<$[Fe/H]$> -1.49 \pm 0.04$
derived from the observed MDF and the inferred global MDF ($<$[Fe/H]$> -1.47$ and $-1.41$ from the pristine and 
extra-gas model, respectively).

Finally, we note that the recovered MDF is valid for the RGB but not necessarily for the entire galaxy. As mentioned earlier, the different time that stars of different mass spend on the RGB introduces a bias in the sense that the observed number of red giants does not accurately reflect the actual relative mass of metals.
The net effect for a metal-poor, mostly old galaxy like Phx is that the MDF inferred from RGB observations is inflated at the high-metallicity end and suppressed at the low-metallicity end compared to the true underlying MDF.
This arises indirectly, because there is a smaller range of stellar mass within the sampled magnitude interval along the RGB for older stars than for younger stars. For the very shallow age-metallicity relation shown by Phoenix, the effect on the MDF is therefore likely to be small, on the order of $10-20\%$ or less per metallicity bin (Ellen Manning, private communication).
A more quantitative estimate requires careful modelling (Manning \& Cole, in prep.), which exceeds the scope of this work.

\subsection{An extremely metal-poor star candidate}

The study of the detailed chemical abundances of extremely metal poor (EMP, [Fe/H] $< -3$) stars
is considered as particularly insightful for understanding the early chemical enrichment of galaxies as well as the properties of the first stars \citep{frebel+norris2015}.

Among the Local Group dwarf galaxies, so far, the detection of EMP stars has mostly been limited to the dwarf galaxies satellites of the MW \citep[e.g.][]{frebel+2010, tafelmeyer+2010, kirby+cohen2012, starkenburg+2013},
being easier to spot in those dwarfs with the lowest mean metallicities. While their number has been significantly increasing in the last few years,
they remain fairly scarce, because searching for them requires either the acquisition of large data-sets or selection methods aimed at isolating
suitable candidates from the overall population followed by high-resolution spectroscopic follow-up \citep[see e.g.][]{simon+2015}. 

Here, we report on an EMP star candidate in our data-set. 
The EMP candidate has a metallicity according to the CaT calibration of $\rm{[Fe/H]}=-3.5\pm0.1$\,dex.
%It is designated with {\it phx12845} in the photometric catalogue and has the following equatorial coordinates: $\alpha_{J2000} = 01^h\,51^m\,01.8^s, \delta_{J2000} = -44^{\circ}\,29'\,17.7''$.   %27.75767900   -44.48826200
We measured a heliocentric velocity of $-16.5\pm5.8$\kms. 
The star is found well within the tidal radius of Phx, about $2.8\arcmin$ from the centre.
%and has a very high probability of membership ($P_{\rm mem}$= 1). 
With a magnitude $V=20.54\pm0.04$\,mag and colour $V-I = 1.31$\,mag, it is among the brightest stars in our sample ($\rm{SNR} = 65$\,px$^{-1}$ very good quality spectrum) and is situated right at the tip of the RGB of Phx. The star can be identified in Fig. \ref{fig:phx-metallicity} as the one with the darkest blue colour.
It has by far the weakest CaT lines in the entire sample with $\rm{EW}_2 = 0.72\pm0.04$\,\AA~and $\rm{EW}_3 = 0.69\pm0.05$\,\AA.
For comparison the second most metal poor star with similar luminosity $V=20.69\pm0.04$\,mag has $\rm{EW}_2 = 1.59\pm0.15$\,\AA~and $\rm{EW}_3 = 1.54\pm0.12$\,\AA~and estimated $\rm{[Fe/H]}=-2.4\pm0.1$\,dex. The EW of the Mg~I line at $8806.8$\,\AA\, is 0.1\AA\,, a value compatible with the object being an EMP star
\citep{battaglia+starkenburg2012}.

We performed spectral synthesis using FER\reflectbox{R}E\footnote{\url{http://vivaldi.ll.iac.es/galeria/callende/ferre/}} \citep{prieto+2006}.
To this end, we used a grid of synthetic spectra computed with ASSET \citep{koesterke+2008} from the ATLAS9 model atmospheres \citep{castelli+kurucz2004}. In this case, we used a grid with five dimensions, [Fe/H], [$\alpha$/Fe], $\log v_t$ (microturbulence velocity), $T_{eff}$ and $\log g$. Other abundances such as C were fixed to the Solar values.
The best fit model predicts [Fe/H]$=-3.89\pm0.15$\,dex, [$\alpha$/Fe]$=0.56\pm0.30$\,dex, $T_{eff}=3696\pm65$\,K, and $\log g=1.7\pm0.4$\,dex.

An independent spectral synthesis of the object, kindly performed by M.~Beasley with pPXF \citep{cappellari2012} with the synthetic spectral models used in \citet{starkenburg+2010}, favours
a slightly higher [Fe/H]$=-3.0$, effective temperature $4700$\,K, [$\alpha$/Fe]$=0.4$ and $logg = 2.2$\,dex.

The three independent methods point to this star being an EMP giant star.
Finally, the Besan\c{c}on model predicts the contamination
to be almost entirely due to MW dwarf stars from the halo and thick disc, hence it is unlikely the object is a giant star belonging to the MW.

We conclude that it is highly likely that we have uncovered the presence of an EMP giant star in Phoenix.
%Note: A few candidate EMP stars are also visible in the Kirby's MDF of dIrrs and dIrr/dSphs

\section{Conclusions} \label{sec:conclusions}
In this work, we presented results from our VLT/FORS2 multi-object spectroscopy of $280$ targets, $196$ of which were identified as very likely RGB stars,
members of the Phoenix dwarf galaxy. These data are used to characterise the
internal, wide-area kinematic and metallicity properties of this system, and represent the first spectroscopic 
metallicity measurements available for this galaxy. We also present a modification to the analytical chemical enrichment models of \citet{kirby+2011} that accounts for metallicity gradients and resulting observational biases.

Phoenix is approaching the MW with a heliocentric (GSR) velocity of $-21.2\pm 1.0$\kms ($-108.6 \pm 1.0$\kms). Our determination confirms
the systemic velocity measurement by \citet{irwin+tolstoy2002} and re-enforces the hypothesis of a physical association between
Phoenix and the nearby HI cloud moving at $-23$\,\kms \citep{germain+1999}.
%The current distance, systemic velocity and presence of HI gas in this small galactic system all point towards Phoenix being at its first approach towards the MW, and therefore having evolved undisturbed by interactions with a large galaxy.

The analysis of the internal kinematic properties reveals the presence of a velocity gradient along the projected minor axis.
Since we can exclude perspective effects as possible causes of the observed gradients,
we deem it as highly likely to have detected prolate rotation in this system. Interestingly,
the rotation signal is aligned with the spatial distribution of young stars, known to be almost perpendicular to the
spatial distribution of Phoenix main body. There is only one other Local Group dwarf galaxy known
to display prolate rotation, And~II, likely caused by a dwarf-dwarf merger or accretion \citep{amorisco+2014}. We speculate that also Phoenix
might have experienced a similar event in the past, giving rise both to the prolate rotation and the peculiar distribution of
young stars.

As in other dwarf galaxies of similar luminosity, the metallicity distribution function of Phx RGB stars spans almost 3\,dex in [Fe/H].
When corrected for the spatial sampling of the spectroscopic data-set, the mean [Fe/H]$=-1.47$\,dex from the resulting MDF would place Phx
in general agreement with the [Fe/H]-L$_V$ relation defined by other Local Group galaxies \citep{kirby+2013}.

We find a clear negative metallicity gradient, which can be approximated by a linear relation with a slope 
0.13$\pm$0.01 dex arcmin$^{-1}$.
Similarly steep gradients have been observed in most MW dwarf spheroidal galaxies of luminosity comparable to Phx.
The fact that such steep metallicity gradients are present in dwarf galaxies inhabiting a dense environment
(i.e. satellites of the MW) and in a system found beyond the MW virial radius and likely at its first approach towards the MW, points towards that the conclusion that
metallicity gradients are an intrinsic property of these low luminosity galaxies.

%In that framework we find central SNe effective yield in Phx of $p_0 = 0.13\pm0.01\,Z/Z_{\odot}$ and successfully recover the observed metallicity gradient.
%We argue that Phx likely did not accrete a large amount of gas throughout its evolution, although our data is not sufficient to conclusively answer this question.

Wide-area spectroscopic surveys of Local Group dwarf galaxies continue unveiling surprising characteristics of these systems
at the low mass end of the galaxy mass function and provide insights into the mechanisms driving the evolution of this
numerous galaxy population.

\section*{Acknowledgements}

We thank the referee Evan Kirby for his positive and constructive report.
We thank Michael Hilker for his help with the data reduction process, 
Mike Beasley for performing the spectral synthesis of the EMP candidate star and Ryan Leaman for
kindly providing the metallicity gradients of MW dSphs and gas-rich objects shown in Fig.~\ref{fig:phx-metallicity}.
We thank Glenn van de Ven, Morgan Fouesneau and Ryan Leaman 
for insightful discussions on the contents of this study. 
This study was partially financially supported by a 2015 ESO DGDF grant.
NK acknowledges financial support from IAC for a three-weeks visit to the institute.
GB gratefully acknowledges support through a Marie-
Curie action Intra European Fellowship, funded by the European Union Seventh Framework Program (FP7/2007-2013)
under Grant agreement number PIEF-GA-2010-274151,  as
well  as  the  financial  support  by  the  Spanish  Ministry  of
Economy  and  Competitiveness  (MINECO)  under  the  Ramon  y  Cajal  Programme  (RYC-2012-11537).  
AC was supported by a fellowship from the Netherlands Research School for Astronomy (NOVA).
This work made extensive use of the NASA Astrophysics Data System bibliographic services.
This research used the facilities of the Canadian Astronomy Data Centre operated by the National Research Council of Canada with the support of the Canadian Space Agency.

%%%%%%%%%%%%%%%%%%%%%%%%%%%%%%%%%%%%%%%%%%%%%%%%%%



\bibliographystyle{mnras}
\bibliography{Phoenix} % if your bibtex file is called example.bib

\begin{thebibliography}{}
\makeatletter
\relax
\def\mn@urlcharsother{\let\do\@makeother \do\$\do\&\do\#\do\^\do\_\do\%\do\~}
\def\mn@doi{\begingroup\mn@urlcharsother \@ifnextchar [ {\mn@doi@}
  {\mn@doi@[]}}
\def\mn@doi@[#1]#2{\def\@tempa{#1}\ifx\@tempa\@empty \href
  {http://dx.doi.org/#2} {doi:#2}\else \href {http://dx.doi.org/#2} {#1}\fi
  \endgroup}
\def\mn@eprint#1#2{\mn@eprint@#1:#2::\@nil}
\def\mn@eprint@arXiv#1{\href {http://arxiv.org/abs/#1} {{\tt arXiv:#1}}}
\def\mn@eprint@dblp#1{\href {http://dblp.uni-trier.de/rec/bibtex/#1.xml}
  {dblp:#1}}
\def\mn@eprint@#1:#2:#3:#4\@nil{\def\@tempa {#1}\def\@tempb {#2}\def\@tempc
  {#3}\ifx \@tempc \@empty \let \@tempc \@tempb \let \@tempb \@tempa \fi \ifx
  \@tempb \@empty \def\@tempb {arXiv}\fi \@ifundefined
  {mn@eprint@\@tempb}{\@tempb:\@tempc}{\expandafter \expandafter \csname
  mn@eprint@\@tempb\endcsname \expandafter{\@tempc}}}

\bibitem[\protect\citeauthoryear{{Allende Prieto}, {Beers}, {Wilhelm},
  {Newberg}, {Rockosi}, {Yanny}  \& {Lee}}{{Allende Prieto}
  et~al.}{2006}]{prieto+2006}
{Allende Prieto} C.,  {Beers} T.~C.,  {Wilhelm} R.,  {Newberg} H.~J.,
  {Rockosi} C.~M.,  {Yanny} B.,   {Lee} Y.~S.,  2006, \mn@doi [\apj]
  {10.1086/498131}, \href {http://adsabs.harvard.edu/abs/2006ApJ...636..804A}
  {636, 804}

\bibitem[\protect\citeauthoryear{{Amorisco}, {Evans}  \& {van de
  Ven}}{{Amorisco} et~al.}{2014}]{amorisco+2014}
{Amorisco} N.~C.,  {Evans} N.~W.,   {van de Ven} G.,  2014, \mn@doi [\nat]
  {10.1038/nature12995}, \href
  {http://adsabs.harvard.edu/abs/2014Natur.507..335A} {507, 335}

\bibitem[\protect\citeauthoryear{{Aparicio} \& {Gallart}}{{Aparicio} \&
  {Gallart}}{2004}]{aparicio+gallart2004}
{Aparicio} A.,  {Gallart} C.,  2004, \mn@doi [\aj] {10.1086/382836}, \href
  {http://adsabs.harvard.edu/abs/2004AJ....128.1465A} {128, 1465}

\bibitem[\protect\citeauthoryear{{Appenzeller} et~al.,}{{Appenzeller}
  et~al.}{1998}]{appenzeller+1998}
{Appenzeller} I.,  et~al., 1998, The Messenger, \href
  {http://esoads.eso.org/abs/1998Msngr..94....1A} {94, 1}

\bibitem[\protect\citeauthoryear{{Armandroff} \& {Da Costa}}{{Armandroff} \&
  {Da Costa}}{1991}]{armandroff+dacosta1991}
{Armandroff} T.~E.,  {Da Costa} G.~S.,  1991, \mn@doi [\aj] {10.1086/115769},
  \href {http://adsabs.harvard.edu/abs/1991AJ....101.1329A} {101, 1329}

\bibitem[\protect\citeauthoryear{{Armandroff} \& {Zinn}}{{Armandroff} \&
  {Zinn}}{1988}]{armandroff+zinn1988}
{Armandroff} T.~E.,  {Zinn} R.,  1988, \mn@doi [\aj] {10.1086/114792}, \href
  {http://adsabs.harvard.edu/abs/1988AJ.....96...92A} {96, 92}

\bibitem[\protect\citeauthoryear{{Battaglia} \& {Starkenburg}}{{Battaglia} \&
  {Starkenburg}}{2012}]{battaglia+starkenburg2012}
{Battaglia} G.,  {Starkenburg} E.,  2012, \mn@doi [\aap]
  {10.1051/0004-6361/201117557}, \href
  {http://adsabs.harvard.edu/abs/2012A%26A...539A.123B} {539, A123}

\bibitem[\protect\citeauthoryear{{Battaglia} et~al.,}{{Battaglia}
  et~al.}{2006}]{battaglia+2006}
{Battaglia} G.,  et~al., 2006, \mn@doi [\aap] {10.1051/0004-6361:20065720},
  \href {http://adsabs.harvard.edu/abs/2006A%26A...459..423B} {459, 423}

\bibitem[\protect\citeauthoryear{{Battaglia}, {Irwin}, {Tolstoy}, {Hill},
  {Helmi}, {Letarte}  \& {Jablonka}}{{Battaglia}
  et~al.}{2008a}]{battaglia+2008}
{Battaglia} G.,  {Irwin} M.,  {Tolstoy} E.,  {Hill} V.,  {Helmi} A.,  {Letarte}
  B.,   {Jablonka} P.,  2008a, \mn@doi [\mnras]
  {10.1111/j.1365-2966.2007.12532.x}, \href
  {http://adsabs.harvard.edu/abs/2008MNRAS.383..183B} {383, 183}

\bibitem[\protect\citeauthoryear{{Battaglia}, {Helmi}, {Tolstoy}, {Irwin},
  {Hill}  \& {Jablonka}}{{Battaglia} et~al.}{2008b}]{battaglia+2008b}
{Battaglia} G.,  {Helmi} A.,  {Tolstoy} E.,  {Irwin} M.,  {Hill} V.,
  {Jablonka} P.,  2008b, \mn@doi [\apjl] {10.1086/590179}, \href
  {http://adsabs.harvard.edu/abs/2008ApJ...681L..13B} {681, L13}

\bibitem[\protect\citeauthoryear{{Battaglia}, {Tolstoy}, {Helmi}, {Irwin},
  {Parisi}, {Hill}  \& {Jablonka}}{{Battaglia} et~al.}{2011}]{battaglia+2011}
{Battaglia} G.,  {Tolstoy} E.,  {Helmi} A.,  {Irwin} M.,  {Parisi} P.,  {Hill}
  V.,   {Jablonka} P.,  2011, \mn@doi [\mnras]
  {10.1111/j.1365-2966.2010.17745.x}, \href
  {http://adsabs.harvard.edu/abs/2011MNRAS.411.1013B} {411, 1013}

\bibitem[\protect\citeauthoryear{{Battaglia}, {Rejkuba}, {Tolstoy}, {Irwin}  \&
  {Beccari}}{{Battaglia} et~al.}{2012}]{battaglia+2012}
{Battaglia} G.,  {Rejkuba} M.,  {Tolstoy} E.,  {Irwin} M.~J.,   {Beccari} G.,
  2012, \mn@doi [\mnras] {10.1111/j.1365-2966.2012.21286.x}, \href
  {http://adsabs.harvard.edu/abs/2012MNRAS.424.1113B} {424, 1113}

\bibitem[\protect\citeauthoryear{{Binney} \& {Tremaine}}{{Binney} \&
  {Tremaine}}{1987}]{binney+tremaine1987}
{Binney} J.,  {Tremaine} S.,  1987, {Galactic dynamics}

\bibitem[\protect\citeauthoryear{{Bovy}, {Hogg}  \& {Rix}}{{Bovy}
  et~al.}{2009}]{bovy+2009}
{Bovy} J.,  {Hogg} D.~W.,   {Rix} H.-W.,  2009, \mn@doi [\apj]
  {10.1088/0004-637X/704/2/1704}, \href
  {http://adsabs.harvard.edu/abs/2009ApJ...704.1704B} {704, 1704}

\bibitem[\protect\citeauthoryear{{Boylan-Kolchin}, {Bullock}, {Sohn}, {Besla}
  \& {van der Marel}}{{Boylan-Kolchin} et~al.}{2013}]{Boylan-Kolchin+2013}
{Boylan-Kolchin} M.,  {Bullock} J.~S.,  {Sohn} S.~T.,  {Besla} G.,   {van der
  Marel} R.~P.,  2013, \mn@doi [\apj] {10.1088/0004-637X/768/2/140}, \href
  {http://adsabs.harvard.edu/abs/2013ApJ...768..140B} {768, 140}

\bibitem[\protect\citeauthoryear{{Cappellari}}{{Cappellari}}{2012}]{cappellari2012}
{Cappellari} M.,  2012, preprint, \href
  {http://adsabs.harvard.edu/abs/2012arXiv1211.7009C} {} (\mn@eprint {arXiv}
  {1211.7009})

\bibitem[\protect\citeauthoryear{{Carrera}}{{Carrera}}{2012}]{carrera2012}
{Carrera} R.,  2012, \mn@doi [\aap] {10.1051/0004-6361/201219625}, \href
  {http://adsabs.harvard.edu/abs/2012A%26A...544A.109C} {544, A109}

\bibitem[\protect\citeauthoryear{{Carrera}, {Gallart}, {Hardy}, {Aparicio}  \&
  {Zinn}}{{Carrera} et~al.}{2008a}]{carrera+2008b}
{Carrera} R.,  {Gallart} C.,  {Hardy} E.,  {Aparicio} A.,   {Zinn} R.,  2008a,
  \mn@doi [\aj] {10.1088/0004-6256/135/3/836}, \href
  {http://adsabs.harvard.edu/abs/2008AJ....135..836C} {135, 836}

\bibitem[\protect\citeauthoryear{{Carrera}, {Gallart}, {Aparicio}, {Costa},
  {M{\'e}ndez}  \& {No{\"e}l}}{{Carrera} et~al.}{2008b}]{carrera+2008a}
{Carrera} R.,  {Gallart} C.,  {Aparicio} A.,  {Costa} E.,  {M{\'e}ndez} R.~A.,
   {No{\"e}l} N.~E.~D.,  2008b, \mn@doi [\aj] {10.1088/0004-6256/136/3/1039},
  \href {http://adsabs.harvard.edu/abs/2008AJ....136.1039C} {136, 1039}

\bibitem[\protect\citeauthoryear{{Carrera}, {Pancino}, {Gallart}  \& {del
  Pino}}{{Carrera} et~al.}{2013}]{carrera+2013}
{Carrera} R.,  {Pancino} E.,  {Gallart} C.,   {del Pino} A.,  2013, \mn@doi
  [\mnras] {10.1093/mnras/stt1126}, \href
  {http://adsabs.harvard.edu/abs/2013MNRAS.434.1681C} {434, 1681}

\bibitem[\protect\citeauthoryear{{Castelli} \& {Kurucz}}{{Castelli} \&
  {Kurucz}}{2004}]{castelli+kurucz2004}
{Castelli} F.,  {Kurucz} R.~L.,  2004, ArXiv Astrophysics e-prints, \href
  {http://adsabs.harvard.edu/abs/2004astro.ph..5087C} {}

\bibitem[\protect\citeauthoryear{{Cayrel}}{{Cayrel}}{1988}]{cayrel1988}
{Cayrel} R.,  1988, in {Cayrel de Strobel} G.,  {Spite} M.,  eds,  IAU
  Symposium Vol. 132, The Impact of Very High S/N Spectroscopy on Stellar
  Physics. p.~345

\bibitem[\protect\citeauthoryear{{Cenarro}, {Cardiel}, {Gorgas}, {Peletier},
  {Vazdekis}  \& {Prada}}{{Cenarro} et~al.}{2001}]{cenarro+2001}
{Cenarro} A.~J.,  {Cardiel} N.,  {Gorgas} J.,  {Peletier} R.~F.,  {Vazdekis}
  A.,   {Prada} F.,  2001, \mn@doi [\mnras] {10.1046/j.1365-8711.2001.04688.x},
  \href {http://adsabs.harvard.edu/abs/2001MNRAS.326..959C} {326, 959}

\bibitem[\protect\citeauthoryear{{Cole}, {Smecker-Hane}, {Tolstoy}, {Bosler}
  \& {Gallagher}}{{Cole} et~al.}{2004}]{cole+2004}
{Cole} A.~A.,  {Smecker-Hane} T.~A.,  {Tolstoy} E.,  {Bosler} T.~L.,
  {Gallagher} J.~S.,  2004, \mn@doi [\mnras]
  {10.1111/j.1365-2966.2004.07223.x}, \href
  {http://adsabs.harvard.edu/abs/2004MNRAS.347..367C} {347, 367}

\bibitem[\protect\citeauthoryear{{Cole}, {Tolstoy}, {Gallagher}  \&
  {Smecker-Hane}}{{Cole} et~al.}{2005}]{cole+2005}
{Cole} A.~A.,  {Tolstoy} E.,  {Gallagher} III J.~S.,   {Smecker-Hane} T.~A.,
  2005, \mn@doi [\aj] {10.1086/428007}, \href
  {http://adsabs.harvard.edu/abs/2005AJ....129.1465C} {129, 1465}

\bibitem[\protect\citeauthoryear{{Cole}, {Grocholski}, {Geisler}, {Sarajedini},
  {Smith}  \& {Tolstoy}}{{Cole} et~al.}{2009}]{cole+2009}
{Cole} A.~A.,  {Grocholski} A.~J.,  {Geisler} D.,  {Sarajedini} A.,  {Smith}
  V.~V.,   {Tolstoy} E.,  2009, in {Van Loon} J.~T.,  {Oliveira} J.~M.,  eds,
  IAU Symposium Vol. 256, The Magellanic System: Stars, Gas, and Galaxies. pp
  263--268, \mn@doi{10.1017/S174392130802855X}

\bibitem[\protect\citeauthoryear{{Coleman}, {Da Costa}, {Bland-Hawthorn}  \&
  {Freeman}}{{Coleman} et~al.}{2005}]{coleman+2005}
{Coleman} M.~G.,  {Da Costa} G.~S.,  {Bland-Hawthorn} J.,   {Freeman} K.~C.,
  2005, \mn@doi [\aj] {10.1086/427966}, \href
  {http://adsabs.harvard.edu/abs/2005AJ....129.1443C} {129, 1443}

\bibitem[\protect\citeauthoryear{{Collins} et~al.,}{{Collins}
  et~al.}{2013}]{collins+2013}
{Collins} M.~L.~M.,  et~al., 2013, \mn@doi [\apj]
  {10.1088/0004-637X/768/2/172}, \href
  {http://adsabs.harvard.edu/abs/2013ApJ...768..172C} {768, 172}

\bibitem[\protect\citeauthoryear{{De Rijcke}, {Dejonghe}, {Zeilinger}  \&
  {Hau}}{{De Rijcke} et~al.}{2004}]{derijcke+2004}
{De Rijcke} S.,  {Dejonghe} H.,  {Zeilinger} W.~W.,   {Hau} G.~K.~T.,  2004,
  \mn@doi [\aap] {10.1051/0004-6361:20041205}, \href
  {http://adsabs.harvard.edu/abs/2004A%26A...426...53D} {426, 53}

\bibitem[\protect\citeauthoryear{{Dehnen} \& {Binney}}{{Dehnen} \&
  {Binney}}{1998}]{dehnen+binney1998}
{Dehnen} W.,  {Binney} J.,  1998, \mn@doi [\mnras]
  {10.1046/j.1365-8711.1998.01282.x}, \href
  {http://adsabs.harvard.edu/abs/1998MNRAS.294..429D} {294, 429}

\bibitem[\protect\citeauthoryear{{Emsellem} et~al.,}{{Emsellem}
  et~al.}{2004}]{emsellem+2004}
{Emsellem} E.,  et~al., 2004, \mn@doi [\mnras]
  {10.1111/j.1365-2966.2004.07948.x}, \href
  {http://adsabs.harvard.edu/abs/2004MNRAS.352..721E} {352, 721}

\bibitem[\protect\citeauthoryear{{Faria}, {Feltzing}, {Lundstr{\"o}m},
  {Gilmore}, {Wahlgren}, {Ardeberg}  \& {Linde}}{{Faria}
  et~al.}{2007}]{faria+2007}
{Faria} D.,  {Feltzing} S.,  {Lundstr{\"o}m} I.,  {Gilmore} G.,  {Wahlgren}
  G.~M.,  {Ardeberg} A.,   {Linde} P.,  2007, \mn@doi [\aap]
  {10.1051/0004-6361:20065244}, \href
  {http://adsabs.harvard.edu/abs/2007A%26A...465..357F} {465, 357}

\bibitem[\protect\citeauthoryear{{Fouquet}, {Lokas}, {del Pino}  \&
  {Ebrova}}{{Fouquet} et~al.}{2016}]{fouquet+2016}
{Fouquet} S.,  {Lokas} E.~L.,  {del Pino} A.,   {Ebrova} I.,  2016, preprint,
  \href {http://adsabs.harvard.edu/abs/2016arXiv160609259F} {} (\mn@eprint
  {arXiv} {1606.09259})

\bibitem[\protect\citeauthoryear{{Fraternali}, {Tolstoy}, {Irwin}  \&
  {Cole}}{{Fraternali} et~al.}{2009}]{fraternali+2009}
{Fraternali} F.,  {Tolstoy} E.,  {Irwin} M.~J.,   {Cole} A.~A.,  2009, \mn@doi
  [\aap] {10.1051/0004-6361/200810830}, \href
  {http://adsabs.harvard.edu/abs/2009A%26A...499..121F} {499, 121}

\bibitem[\protect\citeauthoryear{{Frebel} \& {Norris}}{{Frebel} \&
  {Norris}}{2015}]{frebel+norris2015}
{Frebel} A.,  {Norris} J.~E.,  2015, \mn@doi [\araa]
  {10.1146/annurev-astro-082214-122423}, \href
  {http://adsabs.harvard.edu/abs/2015ARA%26A..53..631F} {53, 631}

\bibitem[\protect\citeauthoryear{{Frebel}, {Simon}, {Geha}  \&
  {Willman}}{{Frebel} et~al.}{2010}]{frebel+2010}
{Frebel} A.,  {Simon} J.~D.,  {Geha} M.,   {Willman} B.,  2010, \mn@doi [\apj]
  {10.1088/0004-637X/708/1/560}, \href
  {http://adsabs.harvard.edu/abs/2010ApJ...708..560F} {708, 560}

\bibitem[\protect\citeauthoryear{{Gallart}, {Mart{\'{\i}}nez-Delgado},
  {G{\'o}mez-Flechoso}  \& {Mateo}}{{Gallart} et~al.}{2001}]{gallart+2001}
{Gallart} C.,  {Mart{\'{\i}}nez-Delgado} D.,  {G{\'o}mez-Flechoso} M.~A.,
  {Mateo} M.,  2001, \mn@doi [\aj] {10.1086/320395}, \href
  {http://adsabs.harvard.edu/abs/2001AJ....121.2572G} {121, 2572}

\bibitem[\protect\citeauthoryear{{Gallart} et~al.,}{{Gallart}
  et~al.}{2015}]{gallart+2015}
{Gallart} C.,  et~al., 2015, \mn@doi [\apjl] {10.1088/2041-8205/811/2/L18},
  \href {http://adsabs.harvard.edu/abs/2015ApJ...811L..18G} {811, L18}

\bibitem[\protect\citeauthoryear{{Gebhardt}, {Pryor}, {O'Connell}, {Williams}
  \& {Hesser}}{{Gebhardt} et~al.}{2000}]{gebhardt+2000}
{Gebhardt} K.,  {Pryor} C.,  {O'Connell} R.~D.,  {Williams} T.~B.,   {Hesser}
  J.~E.,  2000, \mn@doi [\aj] {10.1086/301275}, \href
  {http://adsabs.harvard.edu/abs/2000AJ....119.1268G} {119, 1268}

\bibitem[\protect\citeauthoryear{{Gwinn}, {Moran}  \& {Reid}}{{Gwinn}
  et~al.}{1992}]{gwinn+1992}
{Gwinn} C.~R.,  {Moran} J.~M.,   {Reid} M.~J.,  1992, \mn@doi [\apj]
  {10.1086/171493}, \href {http://adsabs.harvard.edu/abs/1992ApJ...393..149G}
  {393, 149}

\bibitem[\protect\citeauthoryear{{Harris}}{{Harris}}{1996}]{harris1996}
{Harris} W.~E.,  1996, \mn@doi [\aj] {10.1086/118116}, \href
  {http://adsabs.harvard.edu/abs/1996AJ....112.1487H} {112, 1487}

\bibitem[\protect\citeauthoryear{Hastings}{Hastings}{1970}]{hastings70}
Hastings W.~K.,  1970, Biometrika, 57, 97

\bibitem[\protect\citeauthoryear{{Hendricks}, {Koch}, {Walker}, {Johnson},
  {Pe{\~n}arrubia}  \& {Gilmore}}{{Hendricks} et~al.}{2014a}]{hendricks+2014}
{Hendricks} B.,  {Koch} A.,  {Walker} M.,  {Johnson} C.~I.,  {Pe{\~n}arrubia}
  J.,   {Gilmore} G.,  2014a, \mn@doi [\aap] {10.1051/0004-6361/201424645},
  \href {http://adsabs.harvard.edu/abs/2014A%26A...572A..82H} {572, A82}

\bibitem[\protect\citeauthoryear{{Hendricks}, {Koch}, {Lanfranchi}, {Boeche},
  {Walker}, {Johnson}, {Pe{\~n}arrubia}  \& {Gilmore}}{{Hendricks}
  et~al.}{2014b}]{hendricks+2014b}
{Hendricks} B.,  {Koch} A.,  {Lanfranchi} G.~A.,  {Boeche} C.,  {Walker} M.,
  {Johnson} C.~I.,  {Pe{\~n}arrubia} J.,   {Gilmore} G.,  2014b, \mn@doi [\apj]
  {10.1088/0004-637X/785/2/102}, \href
  {http://adsabs.harvard.edu/abs/2014ApJ...785..102H} {785, 102}

\bibitem[\protect\citeauthoryear{{Hidalgo}, {Aparicio},
  {Mart{\'{\i}}nez-Delgado}  \& {Gallart}}{{Hidalgo}
  et~al.}{2009}]{hidalgo+2009}
{Hidalgo} S.~L.,  {Aparicio} A.,  {Mart{\'{\i}}nez-Delgado} D.,   {Gallart} C.,
   2009, \mn@doi [\apj] {10.1088/0004-637X/705/1/704}, \href
  {http://adsabs.harvard.edu/abs/2009ApJ...705..704H} {705, 704}

\bibitem[\protect\citeauthoryear{{Hidalgo} et~al.,}{{Hidalgo}
  et~al.}{2013}]{hidalgo+2013}
{Hidalgo} S.~L.,  et~al., 2013, \mn@doi [\apj] {10.1088/0004-637X/778/2/103},
  \href {http://adsabs.harvard.edu/abs/2013ApJ...778..103H} {778, 103}

\bibitem[\protect\citeauthoryear{{Ho} et~al.,}{{Ho} et~al.}{2012}]{ho+2012}
{Ho} N.,  et~al., 2012, \mn@doi [\apj] {10.1088/0004-637X/758/2/124}, \href
  {http://adsabs.harvard.edu/abs/2012ApJ...758..124H} {758, 124}

\bibitem[\protect\citeauthoryear{{Ho}, {Geha}, {Tollerud}, {Zinn},
  {Guhathakurta}  \& {Vargas}}{{Ho} et~al.}{2015}]{ho+2015}
{Ho} N.,  {Geha} M.,  {Tollerud} E.~J.,  {Zinn} R.,  {Guhathakurta} P.,
  {Vargas} L.~C.,  2015, \mn@doi [\apj] {10.1088/0004-637X/798/2/77}, \href
  {http://adsabs.harvard.edu/abs/2015ApJ...798...77H} {798, 77}

\bibitem[\protect\citeauthoryear{{Holtzman}, {Smith}  \&
  {Grillmair}}{{Holtzman} et~al.}{2000}]{holtzman+2000}
{Holtzman} J.~A.,  {Smith} G.~H.,   {Grillmair} C.,  2000, \mn@doi [\aj]
  {10.1086/316844}, \href {http://adsabs.harvard.edu/abs/2000AJ....120.3060H}
  {120, 3060}

\bibitem[\protect\citeauthoryear{{Irwin} \& {Tolstoy}}{{Irwin} \&
  {Tolstoy}}{2002}]{irwin+tolstoy2002}
{Irwin} M.,  {Tolstoy} E.,  2002, \mn@doi [\mnras]
  {10.1046/j.1365-8711.2002.05802.x}, \href
  {http://adsabs.harvard.edu/abs/2002MNRAS.336..643I} {336, 643}

\bibitem[\protect\citeauthoryear{{Kirby} \& {Cohen}}{{Kirby} \&
  {Cohen}}{2012}]{kirby+cohen2012}
{Kirby} E.~N.,  {Cohen} J.~G.,  2012, \mn@doi [\aj]
  {10.1088/0004-6256/144/6/168}, \href
  {http://adsabs.harvard.edu/abs/2012AJ....144..168K} {144, 168}

\bibitem[\protect\citeauthoryear{{Kirby}, {Lanfranchi}, {Simon}, {Cohen}  \&
  {Guhathakurta}}{{Kirby} et~al.}{2011a}]{kirby+2011}
{Kirby} E.~N.,  {Lanfranchi} G.~A.,  {Simon} J.~D.,  {Cohen} J.~G.,
  {Guhathakurta} P.,  2011a, \mn@doi [\apj] {10.1088/0004-637X/727/2/78}, \href
  {http://adsabs.harvard.edu/abs/2011ApJ...727...78K} {727, 78}

\bibitem[\protect\citeauthoryear{{Kirby}, {Martin}  \& {Finlator}}{{Kirby}
  et~al.}{2011b}]{kirby+2011b}
{Kirby} E.~N.,  {Martin} C.~L.,   {Finlator} K.,  2011b, \mn@doi [\apjl]
  {10.1088/2041-8205/742/2/L25}, \href
  {http://adsabs.harvard.edu/abs/2011ApJ...742L..25K} {742, L25}

\bibitem[\protect\citeauthoryear{{Kirby}, {Cohen}, {Guhathakurta}, {Cheng},
  {Bullock}  \& {Gallazzi}}{{Kirby} et~al.}{2013}]{kirby+2013}
{Kirby} E.~N.,  {Cohen} J.~G.,  {Guhathakurta} P.,  {Cheng} L.,  {Bullock}
  J.~S.,   {Gallazzi} A.,  2013, \mn@doi [\apj] {10.1088/0004-637X/779/2/102},
  \href {http://adsabs.harvard.edu/abs/2013ApJ...779..102K} {779, 102}

\bibitem[\protect\citeauthoryear{{Kirby}, {Bullock}, {Boylan-Kolchin},
  {Kaplinghat}  \& {Cohen}}{{Kirby} et~al.}{2014}]{kirby+2014}
{Kirby} E.~N.,  {Bullock} J.~S.,  {Boylan-Kolchin} M.,  {Kaplinghat} M.,
  {Cohen} J.~G.,  2014, \mn@doi [\mnras] {10.1093/mnras/stu025}, \href
  {http://adsabs.harvard.edu/abs/2014MNRAS.439.1015K} {439, 1015}

\bibitem[\protect\citeauthoryear{{Kleyna}, {Wilkinson}, {Evans}, {Gilmore}  \&
  {Frayn}}{{Kleyna} et~al.}{2002}]{kleyna+2002}
{Kleyna} J.,  {Wilkinson} M.~I.,  {Evans} N.~W.,  {Gilmore} G.,   {Frayn} C.,
  2002, \mn@doi [\mnras] {10.1046/j.1365-8711.2002.05155.x}, \href
  {http://adsabs.harvard.edu/abs/2002MNRAS.330..792K} {330, 792}

\bibitem[\protect\citeauthoryear{{Koch}, {Grebel}, {Wyse}, {Kleyna},
  {Wilkinson}, {Harbeck}, {Gilmore}  \& {Evans}}{{Koch}
  et~al.}{2006}]{koch+2006}
{Koch} A.,  {Grebel} E.~K.,  {Wyse} R.~F.~G.,  {Kleyna} J.~T.,  {Wilkinson}
  M.~I.,  {Harbeck} D.~R.,  {Gilmore} G.~F.,   {Evans} N.~W.,  2006, \mn@doi
  [\aj] {10.1086/499490}, \href
  {http://adsabs.harvard.edu/abs/2006AJ....131..895K} {131, 895}

\bibitem[\protect\citeauthoryear{{Koch}, {Wilkinson}, {Kleyna}, {Gilmore},
  {Grebel}, {Mackey}, {Evans}  \& {Wyse}}{{Koch} et~al.}{2007}]{koch+2007}
{Koch} A.,  {Wilkinson} M.~I.,  {Kleyna} J.~T.,  {Gilmore} G.~F.,  {Grebel}
  E.~K.,  {Mackey} A.~D.,  {Evans} N.~W.,   {Wyse} R.~F.~G.,  2007, \mn@doi
  [\apj] {10.1086/510879}, \href
  {http://adsabs.harvard.edu/abs/2007ApJ...657..241K} {657, 241}

\bibitem[\protect\citeauthoryear{{Koch}, {McWilliam}, {Grebel}, {Zucker}  \&
  {Belokurov}}{{Koch} et~al.}{2008}]{koch+2008}
{Koch} A.,  {McWilliam} A.,  {Grebel} E.~K.,  {Zucker} D.~B.,   {Belokurov} V.,
   2008, \mn@doi [\apjl] {10.1086/595001}, \href
  {http://adsabs.harvard.edu/abs/2008ApJ...688L..13K} {688, L13}

\bibitem[\protect\citeauthoryear{{Koesterke}, {Allende Prieto}  \&
  {Lambert}}{{Koesterke} et~al.}{2008}]{koesterke+2008}
{Koesterke} L.,  {Allende Prieto} C.,   {Lambert} D.~L.,  2008, \mn@doi [\apj]
  {10.1086/587471}, \href {http://adsabs.harvard.edu/abs/2008ApJ...680..764K}
  {680, 764}

\bibitem[\protect\citeauthoryear{{Lane}, {Kiss}, {Lewis}, {Ibata}, {Siebert},
  {Bedding}, {Sz{\'e}kely}  \& {Szab{\'o}}}{{Lane} et~al.}{2011}]{lane+2011}
{Lane} R.~R.,  {Kiss} L.~L.,  {Lewis} G.~F.,  {Ibata} R.~A.,  {Siebert} A.,
  {Bedding} T.~R.,  {Sz{\'e}kely} P.,   {Szab{\'o}} G.~M.,  2011, \mn@doi
  [\aap] {10.1051/0004-6361/201116660}, \href
  {http://adsabs.harvard.edu/abs/2011A%26A...530A..31L} {530, A31}

\bibitem[\protect\citeauthoryear{{Larsen}, {Strader}  \& {Brodie}}{{Larsen}
  et~al.}{2012}]{larsen+2012}
{Larsen} S.~S.,  {Strader} J.,   {Brodie} J.~P.,  2012, \mn@doi [\aap]
  {10.1051/0004-6361/201219897}, \href
  {http://adsabs.harvard.edu/abs/2012A%26A...544L..14L} {544, L14}

\bibitem[\protect\citeauthoryear{{Leaman}, {Cole}, {Venn}, {Tolstoy}, {Irwin},
  {Szeifert}, {Skillman}  \& {McConnachie}}{{Leaman}
  et~al.}{2009}]{leaman+2009}
{Leaman} R.,  {Cole} A.~A.,  {Venn} K.~A.,  {Tolstoy} E.,  {Irwin} M.~J.,
  {Szeifert} T.,  {Skillman} E.~D.,   {McConnachie} A.~W.,  2009, \mn@doi
  [\apj] {10.1088/0004-637X/699/1/1}, \href
  {http://adsabs.harvard.edu/abs/2009ApJ...699....1L} {699, 1}

\bibitem[\protect\citeauthoryear{{Leaman} et~al.,}{{Leaman}
  et~al.}{2013}]{leaman+2013}
{Leaman} R.,  et~al., 2013, \mn@doi [\apj] {10.1088/0004-637X/767/2/131}, \href
  {http://adsabs.harvard.edu/abs/2013ApJ...767..131L} {767, 131}

\bibitem[\protect\citeauthoryear{{Lewis}, {Ibata}, {Chapman}, {McConnachie},
  {Irwin}, {Tolstoy}  \& {Tanvir}}{{Lewis} et~al.}{2007}]{lewis+2007}
{Lewis} G.~F.,  {Ibata} R.~A.,  {Chapman} S.~C.,  {McConnachie} A.,  {Irwin}
  M.~J.,  {Tolstoy} E.,   {Tanvir} N.~R.,  2007, \mn@doi [\mnras]
  {10.1111/j.1365-2966.2007.11395.x}, \href
  {http://adsabs.harvard.edu/abs/2007MNRAS.375.1364L} {375, 1364}

\bibitem[\protect\citeauthoryear{{{\L}okas}, {Ebrov{\'a}}, {Pino}  \&
  {Semczuk}}{{{\L}okas} et~al.}{2014}]{lokas+2014}
{{\L}okas} E.~L.,  {Ebrov{\'a}} I.,  {Pino} A.~d.,   {Semczuk} M.,  2014,
  \mn@doi [\mnras] {10.1093/mnrasl/slu128}, \href
  {http://adsabs.harvard.edu/abs/2014MNRAS.445L...6L} {445, L6}

\bibitem[\protect\citeauthoryear{{Lynden-Bell}}{{Lynden-Bell}}{1975}]{lynden-bell1975}
{Lynden-Bell} D.,  1975, \mn@doi [Vistas in Astronomy]
  {10.1016/0083-6656(75)90005-7}, \href
  {http://adsabs.harvard.edu/abs/1975VA.....19..299L} {19, 299}

\bibitem[\protect\citeauthoryear{{Markwardt}}{{Markwardt}}{2009}]{markwardt2009}
{Markwardt} C.~B.,  2009, in {Bohlender} D.~A.,  {Durand} D.,   {Dowler} P.,
  eds,  Astronomical Society of the Pacific Conference Series Vol. 411,
  Astronomical Data Analysis Software and Systems XVIII. p.~251 (\mn@eprint
  {arXiv} {0902.2850})

\bibitem[\protect\citeauthoryear{{Mart{\'{\i}}nez-Delgado}, {Gallart}  \&
  {Aparicio}}{{Mart{\'{\i}}nez-Delgado} et~al.}{1999a}]{martinez-delgado+1999}
{Mart{\'{\i}}nez-Delgado} D.,  {Gallart} C.,   {Aparicio} A.,  1999a, \mn@doi
  [\aj] {10.1086/300967}, \href
  {http://adsabs.harvard.edu/abs/1999AJ....118..862M} {118, 862}

\bibitem[\protect\citeauthoryear{{Mart{\'{\i}}nez-Delgado}, {Gallart}  \&
  {Aparicio}}{{Mart{\'{\i}}nez-Delgado} et~al.}{1999b}]{MD+1999}
{Mart{\'{\i}}nez-Delgado} D.,  {Gallart} C.,   {Aparicio} A.,  1999b, \mn@doi
  [\aj] {10.1086/300967}, \href
  {http://adsabs.harvard.edu/abs/1999AJ....118..862M} {118, 862}

\bibitem[\protect\citeauthoryear{{Mateo}}{{Mateo}}{1998}]{mateo1998}
{Mateo} M.~L.,  1998, \mn@doi [\araa] {10.1146/annurev.astro.36.1.435}, \href
  {http://adsabs.harvard.edu/abs/1998ARA%26A..36..435M} {36, 435}

\bibitem[\protect\citeauthoryear{{Mayer}, {Mastropietro}, {Wadsley}, {Stadel}
  \& {Moore}}{{Mayer} et~al.}{2006}]{mayer+2006}
{Mayer} L.,  {Mastropietro} C.,  {Wadsley} J.,  {Stadel} J.,   {Moore} B.,
  2006, \mn@doi [\mnras] {10.1111/j.1365-2966.2006.10403.x}, \href
  {http://adsabs.harvard.edu/abs/2006MNRAS.369.1021M} {369, 1021}

\bibitem[\protect\citeauthoryear{{McConnachie}}{{McConnachie}}{2012}]{mcconnachie2012}
{McConnachie} A.~W.,  2012, \mn@doi [\aj] {10.1088/0004-6256/144/1/4}, \href
  {http://adsabs.harvard.edu/abs/2012AJ....144....4M} {144, 4}

\bibitem[\protect\citeauthoryear{{McConnachie} \& {Irwin}}{{McConnachie} \&
  {Irwin}}{2006}]{mcconnachie+irwin2006}
{McConnachie} A.~W.,  {Irwin} M.~J.,  2006, \mn@doi [\mnras]
  {10.1111/j.1365-2966.2005.09806.x}, \href
  {http://adsabs.harvard.edu/abs/2006MNRAS.365.1263M} {365, 1263}

\bibitem[\protect\citeauthoryear{{Mu{\~n}oz} et~al.,}{{Mu{\~n}oz}
  et~al.}{2005}]{munoz+2005}
{Mu{\~n}oz} R.~R.,  et~al., 2005, \mn@doi [\apjl] {10.1086/497396}, \href
  {http://adsabs.harvard.edu/abs/2005ApJ...631L.137M} {631, L137}

\bibitem[\protect\citeauthoryear{{Pietrinferni}, {Cassisi}, {Salaris}  \&
  {Castelli}}{{Pietrinferni} et~al.}{2004}]{pietrinferni+2004}
{Pietrinferni} A.,  {Cassisi} S.,  {Salaris} M.,   {Castelli} F.,  2004,
  \mn@doi [\apj] {10.1086/422498}, \href
  {http://adsabs.harvard.edu/abs/2004ApJ...612..168P} {612, 168}

\bibitem[\protect\citeauthoryear{{Robin}, {Reyl{\'e}}, {Derri{\`e}re}  \&
  {Picaud}}{{Robin} et~al.}{2003}]{robin+2003}
{Robin} A.~C.,  {Reyl{\'e}} C.,  {Derri{\`e}re} S.,   {Picaud} S.,  2003,
  \mn@doi [\aap] {10.1051/0004-6361:20031117}, \href
  {http://adsabs.harvard.edu/abs/2003A%26A...409..523R} {409, 523}

\bibitem[\protect\citeauthoryear{{Rutledge}, {Hesser}  \& {Stetson}}{{Rutledge}
  et~al.}{1997}]{rutledge+1997}
{Rutledge} G.~A.,  {Hesser} J.~E.,   {Stetson} P.~B.,  1997, \mn@doi [\pasp]
  {10.1086/133959}, \href {http://adsabs.harvard.edu/abs/1997PASP..109..907R}
  {109, 907}

\bibitem[\protect\citeauthoryear{{Schechter} \& {Gunn}}{{Schechter} \&
  {Gunn}}{1978}]{schechter+gunn1978}
{Schechter} P.~L.,  {Gunn} J.~E.,  1978, \mn@doi [\aj] {10.1086/112324}, \href
  {http://adsabs.harvard.edu/abs/1978AJ.....83.1360S} {83, 1360}

\bibitem[\protect\citeauthoryear{{Sch{\"o}nrich}, {Binney}  \&
  {Dehnen}}{{Sch{\"o}nrich} et~al.}{2010}]{schoenrich+2010}
{Sch{\"o}nrich} R.,  {Binney} J.,   {Dehnen} W.,  2010, \mn@doi [\mnras]
  {10.1111/j.1365-2966.2010.16253.x}, \href
  {http://adsabs.harvard.edu/abs/2010MNRAS.403.1829S} {403, 1829}

\bibitem[\protect\citeauthoryear{{Schroyen}, {de Rijcke}, {Valcke},
  {Cloet-Osselaer}  \& {Dejonghe}}{{Schroyen} et~al.}{2011}]{schroyen+2011}
{Schroyen} J.,  {de Rijcke} S.,  {Valcke} S.,  {Cloet-Osselaer} A.,
  {Dejonghe} H.,  2011, \mn@doi [\mnras] {10.1111/j.1365-2966.2011.19083.x},
  \href {http://adsabs.harvard.edu/abs/2011MNRAS.416..601S} {416, 601}

\bibitem[\protect\citeauthoryear{{Schroyen}, {De Rijcke}, {Koleva},
  {Cloet-Osselaer}  \& {Vandenbroucke}}{{Schroyen}
  et~al.}{2013}]{schroyen+2013}
{Schroyen} J.,  {De Rijcke} S.,  {Koleva} M.,  {Cloet-Osselaer} A.,
  {Vandenbroucke} B.,  2013, \mn@doi [\mnras] {10.1093/mnras/stt1084}, \href
  {http://adsabs.harvard.edu/abs/2013MNRAS.434..888S} {434, 888}

\bibitem[\protect\citeauthoryear{{Simon}, {Jacobson}, {Frebel}, {Thompson},
  {Adams}  \& {Shectman}}{{Simon} et~al.}{2015}]{simon+2015}
{Simon} J.~D.,  {Jacobson} H.~R.,  {Frebel} A.,  {Thompson} I.~B.,  {Adams}
  J.~J.,   {Shectman} S.~A.,  2015, \mn@doi [\apj]
  {10.1088/0004-637X/802/2/93}, \href
  {http://adsabs.harvard.edu/abs/2015ApJ...802...93S} {802, 93}

\bibitem[\protect\citeauthoryear{{St-Germain}, {Carignan}, {C{\^o}te}  \&
  {Oosterloo}}{{St-Germain} et~al.}{1999}]{germain+1999}
{St-Germain} J.,  {Carignan} C.,  {C{\^o}te} S.,   {Oosterloo} T.,  1999,
  \mn@doi [\aj] {10.1086/301021}, \href
  {http://adsabs.harvard.edu/abs/1999AJ....118.1235S} {118, 1235}

\bibitem[\protect\citeauthoryear{{Starkenburg} et~al.,}{{Starkenburg}
  et~al.}{2010}]{starkenburg+2010}
{Starkenburg} E.,  et~al., 2010, \mn@doi [\aap] {10.1051/0004-6361/200913759},
  \href {http://adsabs.harvard.edu/abs/2010A%26A...513A..34S} {513, A34}

\bibitem[\protect\citeauthoryear{{Starkenburg} et~al.,}{{Starkenburg}
  et~al.}{2013}]{starkenburg+2013}
{Starkenburg} E.,  et~al., 2013, \mn@doi [\aap] {10.1051/0004-6361/201220349},
  \href {http://adsabs.harvard.edu/abs/2013A%26A...549A..88S} {549, A88}

\bibitem[\protect\citeauthoryear{{Strigari}, {Frenk}  \& {White}}{{Strigari}
  et~al.}{2010}]{strigari+2010}
{Strigari} L.~E.,  {Frenk} C.~S.,   {White} S.~D.~M.,  2010, \mn@doi [\mnras]
  {10.1111/j.1365-2966.2010.17287.x}, \href
  {http://adsabs.harvard.edu/abs/2010MNRAS.408.2364S} {408, 2364}

\bibitem[\protect\citeauthoryear{{Swan}, {Cole}, {Tolstoy}  \& {Irwin}}{{Swan}
  et~al.}{2016}]{swan+2016}
{Swan} J.,  {Cole} A.~A.,  {Tolstoy} E.,   {Irwin} M.~J.,  2016, \mn@doi
  [\mnras] {10.1093/mnras/stv2774}, \href
  {http://adsabs.harvard.edu/abs/2016MNRAS.456.4315S} {456, 4315}

\bibitem[\protect\citeauthoryear{{Tafelmeyer} et~al.,}{{Tafelmeyer}
  et~al.}{2010}]{tafelmeyer+2010}
{Tafelmeyer} M.,  et~al., 2010, \mn@doi [\aap] {10.1051/0004-6361/201014733},
  \href {http://adsabs.harvard.edu/abs/2010A%26A...524A..58T} {524, A58}

\bibitem[\protect\citeauthoryear{{Tolstoy}, {Irwin}, {Cole}, {Pasquini},
  {Gilmozzi}  \& {Gallagher}}{{Tolstoy} et~al.}{2001}]{tolstoy+2001}
{Tolstoy} E.,  {Irwin} M.~J.,  {Cole} A.~A.,  {Pasquini} L.,  {Gilmozzi} R.,
  {Gallagher} J.~S.,  2001, \mn@doi [\mnras]
  {10.1046/j.1365-8711.2001.04785.x}, \href
  {http://adsabs.harvard.edu/abs/2001MNRAS.327..918T} {327, 918}

\bibitem[\protect\citeauthoryear{{Tolstoy} et~al.,}{{Tolstoy}
  et~al.}{2004}]{tolstoy+2004}
{Tolstoy} E.,  et~al., 2004, \mn@doi [\apjl] {10.1086/427388}, \href
  {http://adsabs.harvard.edu/abs/2004ApJ...617L.119T} {617, L119}

\bibitem[\protect\citeauthoryear{{Tolstoy}, {Hill}  \& {Tosi}}{{Tolstoy}
  et~al.}{2009}]{tolstoy+2009}
{Tolstoy} E.,  {Hill} V.,   {Tosi} M.,  2009, \mn@doi [\araa]
  {10.1146/annurev-astro-082708-101650}, \href
  {http://adsabs.harvard.edu/abs/2009ARA%26A..47..371T} {47, 371}

\bibitem[\protect\citeauthoryear{{Tonry} \& {Davis}}{{Tonry} \&
  {Davis}}{1979}]{tonry+davis1979}
{Tonry} J.,  {Davis} M.,  1979, \mn@doi [\aj] {10.1086/112569}, \href
  {http://adsabs.harvard.edu/abs/1979AJ.....84.1511T} {84, 1511}

\bibitem[\protect\citeauthoryear{{Tremaine}}{{Tremaine}}{1981}]{tremaine1981}
{Tremaine} S.,  1981, in {Fall} S.~M.,  {Lynden-Bell} D.,  eds, Structure and
  Evolution of Normal Galaxies. pp 67--84

\bibitem[\protect\citeauthoryear{{Tully}, {Shaya}, {Karachentsev}, {Courtois},
  {Kocevski}, {Rizzi}  \& {Peel}}{{Tully} et~al.}{2008}]{tully+2008}
{Tully} R.~B.,  {Shaya} E.~J.,  {Karachentsev} I.~D.,  {Courtois} H.~M.,
  {Kocevski} D.~D.,  {Rizzi} L.,   {Peel} A.,  2008, \mn@doi [\apj]
  {10.1086/527428}, \href {http://adsabs.harvard.edu/abs/2008ApJ...676..184T}
  {676, 184}

\bibitem[\protect\citeauthoryear{{V{\'a}squez}, {Zoccali}, {Hill}, {Gonzalez},
  {Saviane}, {Rejkuba}  \& {Battaglia}}{{V{\'a}squez}
  et~al.}{2015}]{vasquez+2015}
{V{\'a}squez} S.,  {Zoccali} M.,  {Hill} V.,  {Gonzalez} O.~A.,  {Saviane} I.,
  {Rejkuba} M.,   {Battaglia} G.,  2015, \mn@doi [\aap]
  {10.1051/0004-6361/201526534}, \href
  {http://adsabs.harvard.edu/abs/2015A26A...580A.121V} {580, A121}

\bibitem[\protect\citeauthoryear{{Walker}, {Mateo}, {Olszewski}, {Bernstein},
  {Wang}  \& {Woodroofe}}{{Walker} et~al.}{2006}]{walker+2006}
{Walker} M.~G.,  {Mateo} M.,  {Olszewski} E.~W.,  {Bernstein} R.,  {Wang} X.,
  {Woodroofe} M.,  2006, \mn@doi [\aj] {10.1086/500193}, \href
  {http://adsabs.harvard.edu/abs/2006AJ....131.2114W} {131, 2114}

\bibitem[\protect\citeauthoryear{{Walker}, {Mateo}, {Olszewski}, {Sen}  \&
  {Woodroofe}}{{Walker} et~al.}{2009}]{walker+2009}
{Walker} M.~G.,  {Mateo} M.,  {Olszewski} E.~W.,  {Sen} B.,   {Woodroofe} M.,
  2009, \mn@doi [\aj] {10.1088/0004-6256/137/2/3109}, \href
  {http://adsabs.harvard.edu/abs/2009AJ....137.3109W} {137, 3109}

\bibitem[\protect\citeauthoryear{{Walker}, {Olszewski}  \& {Mateo}}{{Walker}
  et~al.}{2015}]{walker+2015}
{Walker} M.~G.,  {Olszewski} E.~W.,   {Mateo} M.,  2015, \mn@doi [\mnras]
  {10.1093/mnras/stv099}, \href
  {http://adsabs.harvard.edu/abs/2015MNRAS.448.2717W} {448, 2717}

\bibitem[\protect\citeauthoryear{{Warren} \& {Cole}}{{Warren} \&
  {Cole}}{2009}]{warren+cole2009}
{Warren} S.~R.,  {Cole} A.~A.,  2009, \mn@doi [\mnras]
  {10.1111/j.1365-2966.2008.14268.x}, \href
  {http://adsabs.harvard.edu/abs/2009MNRAS.393..272W} {393, 272}

\bibitem[\protect\citeauthoryear{{Wheeler}, {Pace}, {Bullock},
  {Boylan-Kolchin}, {Onorbe}, {Fitts}, {Hopkins}  \& {Keres}}{{Wheeler}
  et~al.}{2015}]{wheeler+2015}
{Wheeler} C.,  {Pace} A.~B.,  {Bullock} J.~S.,  {Boylan-Kolchin} M.,  {Onorbe}
  J.,  {Fitts} A.,  {Hopkins} P.~F.,   {Keres} D.,  2015, preprint, \href
  {http://adsabs.harvard.edu/abs/2015arXiv151101095W} {} (\mn@eprint {arXiv}
  {1511.01095})

\bibitem[\protect\citeauthoryear{{Wilkinson}, {Kleyna}, {Evans}, {Gilmore},
  {Irwin}  \& {Grebel}}{{Wilkinson} et~al.}{2004}]{wilkinson+2004}
{Wilkinson} M.~I.,  {Kleyna} J.~T.,  {Evans} N.~W.,  {Gilmore} G.~F.,  {Irwin}
  M.~J.,   {Grebel} E.~K.,  2004, \mn@doi [\apjl] {10.1086/423619}, \href
  {http://adsabs.harvard.edu/abs/2004ApJ...611L..21W} {611, L21}

\bibitem[\protect\citeauthoryear{{Young} \& {Lo}}{{Young} \&
  {Lo}}{1997}]{young+lo1997}
{Young} L.~M.,  {Lo} K.~Y.,  1997, \apj, \href
  {http://adsabs.harvard.edu/abs/1997ApJ...490..710Y} {490, 710}

\bibitem[\protect\citeauthoryear{{Young}, {Skillman}, {Weisz}  \&
  {Dolphin}}{{Young} et~al.}{2007}]{young+2007}
{Young} L.~M.,  {Skillman} E.~D.,  {Weisz} D.~R.,   {Dolphin} A.~E.,  2007,
  \mn@doi [\apj] {10.1086/512153}, \href
  {http://adsabs.harvard.edu/abs/2007ApJ...659..331Y} {659, 331}

\bibitem[\protect\citeauthoryear{{Zaggia}, {Held}, {Sommariva}, {Momany},
  {Saviane}  \& {Rizzi}}{{Zaggia} et~al.}{2011}]{zaggia+2011}
{Zaggia} S.,  {Held} E.~V.,  {Sommariva} V.,  {Momany} Y.,  {Saviane} I.,
  {Rizzi} L.,  2011, in {Koleva} M.,  {Prugniel} P.,   {Vauglin} I.,  eds,  EAS
  Publications Series Vol. 48, EAS Publications Series. pp 215--216,
  \mn@doi{10.1051/eas/1148049}

\bibitem[\protect\citeauthoryear{{de Boer} et~al.,}{{de Boer}
  et~al.}{2012a}]{deboer+2012a}
{de Boer} T.~J.~L.,  et~al., 2012a, \mn@doi [\aap]
  {10.1051/0004-6361/201118378}, \href
  {http://adsabs.harvard.edu/abs/2012A&A...539A.103D} {539, A103}

\bibitem[\protect\citeauthoryear{{de Boer} et~al.,}{{de Boer}
  et~al.}{2012b}]{deboer+2012b}
{de Boer} T.~J.~L.,  et~al., 2012b, \mn@doi [\aap]
  {10.1051/0004-6361/201219547}, \href
  {http://adsabs.harvard.edu/abs/2012A%26A...544A..73D} {544, A73}

\bibitem[\protect\citeauthoryear{{van Dokkum}}{{van
  Dokkum}}{2001}]{vanDokkum2001}
{van Dokkum} P.~G.,  2001, \mn@doi [\pasp] {10.1086/323894}, \href
  {http://adsabs.harvard.edu/abs/2001PASP..113.1420V} {113, 1420}

\bibitem[\protect\citeauthoryear{{van de Rydt}, {Demers}  \& {Kunkel}}{{van de
  Rydt} et~al.}{1991}]{vanderydt+1991}
{van de Rydt} F.,  {Demers} S.,   {Kunkel} W.~E.,  1991, \mn@doi [\aj]
  {10.1086/115861}, \href {http://adsabs.harvard.edu/abs/1991AJ....102..130V}
  {102, 130}

\bibitem[\protect\citeauthoryear{{van de Ven}, {van den Bosch}, {Verolme}  \&
  {de Zeeuw}}{{van de Ven} et~al.}{2006}]{vandeven+2006}
{van de Ven} G.,  {van den Bosch} R.~C.~E.,  {Verolme} E.~K.,   {de Zeeuw}
  P.~T.,  2006, \mn@doi [\aap] {10.1051/0004-6361:20053061}, \href
  {http://adsabs.harvard.edu/abs/2006A%26A...445..513V} {445, 513}

\bibitem[\protect\citeauthoryear{{van den Bosch}, {van de Ven}, {Verolme},
  {Cappellari}  \& {de Zeeuw}}{{van den Bosch} et~al.}{2008}]{vandenbosch+2008}
{van den Bosch} R.~C.~E.,  {van de Ven} G.,  {Verolme} E.~K.,  {Cappellari} M.,
    {de Zeeuw} P.~T.,  2008, \mn@doi [\mnras]
  {10.1111/j.1365-2966.2008.12874.x}, \href
  {http://adsabs.harvard.edu/abs/2008MNRAS.385..647V} {385, 647}

\makeatother
\end{thebibliography}

%\begin{thebibliography}{99}
%uthor A.~N., 2013, Journal of Improbable Astronomy, 1, 1
%bibitem[\protect\citeauthoryear{Others}{2013}]{Others2013}
%\end{thebibliography}

%%%%%%%%%%%%%%%%%%%%%%%%%%%%%%%%%%%%%%%%%%%%%%%%%%

%%%%%%%%%%%%%%%%% APPENDICES %%%%%%%%%%%%%%%%%%%%%

%\appendix

%\section{Extra gas model large sample}

%\begin{figure*}
%	\includegraphics[width=14cm]{extra_gas_large_sample}
%    \caption{Same as Figure \ref{fig:extra-gas} but using 2000 stars instead of 200.
%    }
%    \label{fig:extra-gas-large}
%\end{figure*}


%%%%%%%%%%%%%%%%%%%%%%%%%%%%%%%%%%%%%%%%%%%%%%%%%%

% Don't change these lines
\bsp	% typesetting comment
\label{lastpage}
\end{document}